\documentclass[epj]{svjour}
\usepackage{graphics}
\newcommand{\mod}[1]{\left\vert{#1}\right\vert}
\newcommand{\sitea}{\vec{a}}
\newcommand{\siteb}{\vec{b}}
\newcommand{\sitec}{\vec{c}}
\newcommand{\sited}{\vec{d}}
\newcommand{\sitei}{\vec{i}}
\newcommand{\sitej}{\vec{j}}
\newcommand{\sitek}{\vec{k}}
\newcommand{\siteq}{\vec{q}}
\newcommand{\siteql}{\vec{q}_l}
\newcommand{\siteqn}{\vec{q}_n}
\newcommand{\siteqj}{\vec{q}_j}
\newcommand{\siteqi}{\vec{q}_i}
\newcommand{\sitezero}{\vec{0}}
\newcommand{\I}{{\rm i}}
\newcommand{\diff}{{\rm d}}
\newcommand{\sgn}{{\rm sgn}}
\newcommand{\half}{\frac{1}{2}}

\begin{document}
\title{Analytical and Numerical Treatment of
the Mott--Hubbard Insulator in Infinite Dimensions}
\authorrunning{M.P.~Eastwood et al.}
\titlerunning{Mott--Hubbard Insulator in Infinite Dimensions}

\author{Michael P.~Eastwood \inst{1} 
\and 
Florian Gebhard \inst{2}
\and 
Eva Kalinowski \inst{2}
\and
Satoshi Nishimoto \inst{2}
\and
Reinhard M.~Noack \inst{3}
}                     
%
%
\institute{Department 
of Chemistry and Biochemistry, University of California,
La Jolla, CA 92093-0371, USA
\and 
Fachbereich Physik, Philipps-Universit\"at Marburg,
D-35032 Marburg, Germany
\and 
Institut f\"ur Theoretische Physik~III, Universit\"at Stuttgart,
D-70550 Stuttgart, Germany
}
\date{Received: \today}
%
\abstract{We calculate the density of states in the half-filled 
Hubbard model on a Bethe lattice with infinite connectivity.
Based on our analytical results to second order in $t/U$,
we propose a new `Fixed-Energy Exact Diagonalization' scheme
for the numerical study of the Dynamical Mean-Field Theory.
Corroborated by results from the Random Dispersion Approximation,
we find that the gap opens at $U_{\rm c}=4.43 \pm 0.05$.
Moreover, the density of states near the gap increases algebraically
as a function of frequency with an exponent $\alpha=1/2$ in the
insulating phase. We critically examine other analytical
and numerical approaches 
and specify their merits and limitations 
when applied to the Mott--Hubbard insulator.
\PACS{{71.10Fd}{Lattice fermion models (Hubbard model, etc.)}   
      \and
      {71.27.+a}{Strongly correlated electron systems; heavy fermions} 
      \and
      {71.30+h}{Metal-insulator transitions and other electronic transitions}
     } 
} 
\maketitle
\section{Introduction}
\label{intro}

The Mott--Hubbard metal-insulator transition is a fascinating but
very difficult problem in condensed matter 
theory~\cite{Mottbook,Gebhardbook,RMP,Imada}.
One hopes that the basic mechanism can be understood 
with the help of conceptually simple models such as
the Hubbard Hamiltonian. This model describes
spin-1/2 electrons which 
move on a lattice (band-width $W=4t$) 
and interact locally with strength~$U$. 
For small interactions, $U\ll W$, the model describes
a metal with a finite density of states at the Fermi energy.
For large interactions, $U\gg W$, and 
when there is on average one electron per lattice site
(half band-filling), the model describes an insulator because
it takes an energy of the order of $\Delta(U\gg W) 
\approx U-{\cal O}(W)>0$
to generate a charge excitation, i.e., the single-particle gap $\Delta(U\gg W)$
is finite, irrespective of any symmetry 
breaking~\cite{Mottbook,Gebhardbook}.

These basic insights leave much room for speculations
on the properties of the transition itself, for example,
on the precise value of the critical interaction strength $U_{\rm c}$
at which the gap opens. Because the Mott--Hubbard transition
is a zero-temperature quantum phase transition~\cite{Gebhardbook}, 
it is difficult to access reliably using
analytical or numerical techniques; 
only exact solutions for all interaction strengths can provide
conclusive answers. Unfortunately, Hubbard-type models can be solved
exactly only in one dimension~\cite{LiebWu,GebhardRuckenstein},
and several features in one dimension are not generic to
three-dimensional systems.

The phase diagram of a model in three dimensions
can often be understood from the limit of infinite dimensions.
For correlated lattice electrons~\cite{MV,MHa}, this has led to the
formulation of effective single-impurity models 
whose dynamics must be solved 
self-consistently (Dynamical Mean-Field Theory, 
DMFT)~\cite{RMP,BrandtMielsch,Jarrell}.
An alternative approach, the Random Dispersion Approximation 
(RDA)~\cite{Gebhardbook,RDA}, 
also becomes exact in the limit of infinite dimensions.
However, an exact solution of the DMFT is not feasible
because single-impurity models with a Hubbard interaction
cannot be solved in general.

Using various kinds of approximations, two different scenarios for the
Mott--Hubbard transition have been proposed: 
\begin{description}
\item[\emph{Discontinuous Transition.}] \mbox{} \hfill

The gap jumps to a finite value
when the density of states at the Fermi energy becomes zero at some critical
interaction strength~$U_{{\rm c},2}$; the gap is preformed
above $U_{{\rm c},1}<U_{{\rm c},2}$, and the
co-existing insulating phase is higher in energy 
than the metallic state.
\item[\emph{Continuous Transition.}] \mbox{} \hfill

The gap opens continuously
when the density of states at the Fermi energy becomes zero,
$U_{{\rm c},1}=U_{{\rm c},2}\equiv U_{\rm c}$.
\end{description}
Various approaches to the Mott--Hubbard transition 
for lattices with infinite coordination number
yield results in favor of one or the other 
of the conflicting scenarios; for a review,
see Refs.~\cite{Gebhardbook,RMP}.
Refs.~\cite{RDA}--\cite{Ono} give more recent treatments.
Insightful physical arguments have helped to sharpen the 
physical and mathematical implications 
of the scenario of a preformed gap
transition~\cite{KotliarFisher}--\cite{Kotliarreply} but 
have not been able to 
resolve the issue.

All analytical approaches~\cite{BullaLDMFT}--\cite{LMA}
are necessarily approximate in nature, and
numerical investigations~\cite{7Schwaben}--\cite{andere}
of the dynamical mean-field
equations
involve (i)~discretization, (ii)~numerical diagonalization,
(iii)~interpolation, (iv)~iteration of the self-consistency cycle,
and (v)~extrapolation to the thermodynamic limit.
The Random Dispersion Approximation~\cite{RDA} requires
similar steps in its numerical implementation, apart from step~(iv). 
Therefore, the region of applicability of the available numerical
techniques is not a priori clear either.

In order to address this issue,
two of us (E.K.\ 
and F.G.) have
developed a strong-coupling perturbation theory
at zero temperature~\cite{Eva} which provides a benchmark
test for the available analytical and
numerical techniques for the Mott--Hubbard insulator.
We assess the quality of the Iterated Perturbation Theory (IPT)
and the Local Moment Approach (LMA) as well as
three numerical methods for the Hubbard model in infinite dimensions:
the Random Dispersion Approximation and two
Exact Diagonalization (ED) schemes for the 
Dynamical Mean-Field Theory. As we shall show, this comparison
will provide insight as to the limitations of the various
methods and will motivate an improved ED scheme, the
`Fixed-Energy Exact Diagonalization' (FE-ED).
This scheme successfully reproduces our findings from
the $t/U$ expansion even down to the collapse
of the Mott--Hubbard insulator.

In this work, after the introduction of some definitions 
in Sect.~\ref{sec:Def},
we reanalyze our strong-coupling perturbation theory to second order
in $t/U$ for the Bethe lattice with infinite connectivity.
In Sect.~\ref{sec:Strong} we estimate that
the Mott--Hubbard insulator is destroyed at 
$U_{\rm c}^{\rm sc}/t=4.40 \pm 0.09$.
In Sect.~\ref{sec:FEED} we 
set up the Fixed-Energy Exact Diagonalization
scheme for the Dynamical  Mean-Field Theory.
The FE-ED very well reproduces our analytical findings,
and confirms our conjecture for the critical
interaction strength, $U_{\rm c}^{\rm FE-ED}/t= 4.43 \pm 0.05$.
In Sect.~\ref{sec:RDA} we favorably compare our results with those from
the Random Dispersion Approximation (RDA) which provides
an independent check on our results from the strong-coupling
expansion and the FE-ED.
Therefore, we are confident that we accurately describe
the Mott--Hubbard insulator for all interaction strengths.

In Sect.~\ref{sec:companalyt} we compare our results
from Sects.~\ref{sec:Strong} and~\ref{sec:FEED}
with two analytical
approximations for the Mott--Hubbard insulator, the Iterated Perturbation
Theory (IPT) and the Local Moment Approach (LMA). Both of them
provide a reasonable description down to $U=6 t$ (IPT) and
$U=5.5 t$ (LMA), respectively, where the LMA is generally superior to IPT.
However, both approximations
underestimate the stability of the Mott--Hubbard insulator,
$U_{{\rm c},1}^{\rm IPT}=5.2 t$ and
$U_{{\rm c},1}^{\rm LMA}=4.8 t$.

In Sect.~\ref{sec:garbagenumerics}, we critically
examine two numerical approaches. 
For finite systems, the Caffarel--Krauth 
exact-diagonalization scheme~\cite{Krauth,CK,Ono}
does not provide unique solutions of the self-consistency
equations. Therefore, neither the size of the gap nor
the shape of the density of states can be extrapolated reliably.
The exact diagonalization of the
effective single-impurity Anderson model
in `two-chain geometry'~\cite{RMP,Rozenbergetal,andere} 
fails because the reconstruction of the density of states from its moments
is a numerically too delicate inverse problem.
Conclusions, Sect.~\ref{sec:conclusions}, and appendices on the IPT
and LMA algorithms in the limit of strong coupling
close our work.

\section{Definitions}
\label{sec:Def}

In this section, we discuss the basic properties of 
lattice electrons in the limit of 
infinite dimensions, and we define the Hubbard Hamiltonian, 
the single-particle Green function,
and the corresponding density of states.

\subsection{Hamilton Operator}
\label{subsec:Hamilt}


We investigate spin-1/2 electrons on a lattice whose 
motion is described by the kinetic energy operator
\begin{equation}
\hat{T} =
\sum_{\sitei,\sitej;\sigma} t_{\sitei,\sitej} 
\hat{c}_{\sitei,\sigma}^+\hat{c}_{\sitej,\sigma} \; ,
\end{equation}
where $\hat{c}^+_{\sitei,\sigma}$,
$\hat{c}_{\sitei,\sigma}$ are creation and annihilation operators for
electrons with spin~$\sigma=\uparrow,\downarrow$ on site~$\sitei$.
The matrix elements
$t_{\sitei,\sitej}$ are the electron transfer amplitudes between
sites $\sitei$ and $\sitej$, and $t_{\sitei,\sitei}=0$.
Since we are interested
in the Mott insulating phase, we consider exclusively a half-filled band
where the number of electrons~$N$ equals the
number of lattice sites~$L$. 

For lattices with translational symmetry, $t_{\sitei,\sitej}=
t(\sitei-\sitej)$ and 
the operator for the kinetic energy is diagonal in momentum space,
\begin{equation}
\hat{T} = \sum_{\sitek;\sigma} \epsilon(\sitek) 
\hat{c}^+_{\sitek,\sigma}\hat{c}_{\sitek,\sigma}
\; ,
\end{equation}
where
\begin{equation}
\epsilon(\sitek) = \frac{1}{L} \sum_{\sitei,\sitej} t(\sitei-\sitej)
e^{-\I (\sitei-\sitej) \sitek} \; . 
\end{equation}
The density of states for non-interacting electrons is then given by
\begin{equation}
\rho(\epsilon)= \frac{1}{L} \sum_{\sitek} \delta(\epsilon-\epsilon(\sitek))
 \nonumber \;.
\end{equation}
The $m$-th moment of the density of states is defined by
\begin{equation}
\overline{\epsilon^m} = \int_{-\infty}^{\infty} 
\diff\epsilon\,  \epsilon^m \rho(\epsilon) \;,
\label{energymoments}
\end{equation}
and $\overline{\epsilon}=t(\sitezero)=0$.

In the limit of infinite lattice dimensions and for translationally
invariant systems without nesting, 
the Hubbard model is characterized by $\rho(\epsilon)$ alone, i.e.,
higher-order correlation functions 
in momentum space factorize~\cite{PvDetal}, for example,
\begin{eqnarray}
\rho_{12}(\epsilon_1,\epsilon_2)
&\equiv&
\frac{1}{L} \sum_{\sitek} \delta(\epsilon_2-\epsilon(\sitek+\siteq_2))
\delta(\epsilon_1-\epsilon(\sitek+\siteq_1))
\nonumber \\
&=&
\delta_{\siteq_1,\siteq_2} 
\delta(\epsilon_1-\epsilon_2)
\rho(\epsilon_1) 
\label{RDADq} \\
&& \vphantom{A^X} + (1-\delta_{\siteq_1,\siteq_2}) 
\rho(\epsilon_1)\rho(\epsilon_2) \; .
\nonumber 
\end{eqnarray}
This observation is the basis for the Random Dispersion Approximation
(RDA) which becomes exact in
infinite dimensions for paramagnetic systems, i.e., when
nesting is ignored~\cite{Gebhardbook,RDA}; see Sect.~\ref{sec:RDA}.

For our explicit calculations we shall later use 
the semi-circular density of states
\begin{eqnarray}
\rho_0(\omega)&=& \frac{2}{\pi W}\sqrt{4 -\left(\frac{4\omega}{W}\right)^2\,}
\quad , \quad   (|\omega|\leq W/2) \; ,
\label{rhozero}\\
1 & = & \int_{-W/2}^{W/2} \diff \omega \rho_0(\omega) \; ,
\end{eqnarray}
where $W=4t$ is the band-width. In the following, we take $t\equiv 1$
as our unit of energy.
This density of states is realized for non-interacting
tight-binding electrons on a Bethe lattice of connectivity 
$Z\to\infty$~\cite{Economou}.
Specifically, each site is connected to $Z$~neighbors 
without generating closed loops,
and the electron transfer is restricted to nearest-neighbors:
$t_{\sitei,\sitej}=-t/\sqrt{Z}$ when $\sitei$~and $\sitej$~are
nearest neighbors and
zero otherwise. The limit $Z\to\infty$ is implicitly understood henceforth.
Later on we shall use either of the two equivalent viewpoints,
RDA or tight-binding Bethe lattice, whichever is more convenient for
our considerations.


The electrons are taken to interact only locally,
and the Hubbard interaction reads
\begin{equation}
U \hat{D} = U \sum_{\sitei} \left(\hat{n}_{\sitei,\uparrow}-\frac{1}{2}\right)
\left(\hat{n}_{\sitei,\downarrow}-\frac{1}{2}\right) \; ,
\end{equation}
where $\hat{n}_{\sitei,\sigma}=
\hat{c}^+_{\sitei,\sigma}\hat{c}_{\sitei,\sigma}$ 
is the local density operator at site~$\sitei$ for 
spin~$\sigma$. This 
leads us to the Hubbard model~\cite{HubbardI}, 
\begin{equation}
\hat{H}=\hat{T} + U \hat{D} \; .
\label{generalH}
\end{equation}
The Hamiltonian explicitly exhibits particle-hole symmetry,
i.e., $\hat{H}$ is invariant under the particle-hole transformation 
\begin{equation}
\hat{c}_{\sitei,\sigma}^+ \mapsto (-1)^{|\sitei|}
\hat{c}_{\sitei,\sigma}\quad ; \quad
\hat{c}_{\sitei,\sigma} \mapsto (-1)^{|\sitei|} \hat{c}_{\sitei,\sigma}^+ \; ,
\label{phdef}
\end{equation}
where $|{\sitei}|$ counts the number of nearest-neighbor steps
from the origin of the Bethe lattice to site~$\sitei$.
The chemical potential $\mu=0$ then guarantees a half-filled band
for any temperature~\cite{Gebhardbook}.

For later use we also define the operators for the local electron density,
$\hat{n}_{\sitej}=\hat{n}_{\sitej,\uparrow}+\hat{n}_{\sitej,\downarrow}$.
The operators $\hat{S}_{\sitej}^x=
(\hat{c}^+_{\sitej,\uparrow}\hat{c}_{\sitej,\downarrow}
+ \hat{c}^+_{\sitej,\downarrow}\hat{c}_{\sitej,\uparrow})/2$, 
$\hat{S}_{\sitej}^y=
(\hat{c}^+_{\sitej,\uparrow}\hat{c}_{\sitej,\downarrow}
- \hat{c}^+_{\sitej,\downarrow}\hat{c}_{\sitej,\uparrow})/(2\I)$, 
and $\hat{S}_{\sitej}^z=
(\hat{n}_{\sitej,\uparrow}-\hat{n}_{\sitej,\downarrow})/2$
are the three components of the spin-1/2 vector operator
$\hat{\vec{S}}_{\sitej}$. The operator for the total spin 
$\hat{\vec{S}}=\sum_{\sitej} \hat{\vec{S}}_{\sitej}$ 
commutes with the Hamiltonian.

\subsection{Green Functions}
\label{subsec:Green}

The time-dependent local single-particle Green function at zero temperature
is given by~\cite{Fetter}
\begin{equation}
G(t) = -\I \frac{1}{L} \sum_{\sitei,\sigma} 
\langle \hat{\cal T} [ 
\hat{c}_{\sitei,\sigma}(t)\hat{c}_{\sitei,\sigma}^+] \rangle \; .
\label{GFdef}
\end{equation}
Here $\hat{\cal T}$ is the time-ordering operator and $\langle \ldots \rangle$
implies the average over all ground states with energy $E_0$,
and (taking $\hbar \equiv 1$ henceforth)
\begin{equation}
\hat{c}_{\sitei,\sigma}(t) = \exp(\I \hat{H} t) \hat{c}_{\sitei,\sigma}
                          \exp(-\I \hat{H} t) 
\end{equation}
is the annihilation operator in the Heisenberg picture. 

In the insulating phase we can readily identify 
the contributions from the lower (LHB)
and upper (UHB) Hubbard bands to the 
Fourier transform of the local Green function ($\eta=0^+$),
\begin{eqnarray}
G(\omega) &=& \int_{-\infty}^{\infty} \diff t\,  e^{\I \omega t} G(t) 
= G_{\rm LHB}(\omega ) + G_{\rm UHB}(\omega ) \; , \nonumber \\
G_{\rm LHB}(\omega ) &=&  \frac{1}{L} \sum_{\sitei,\sigma}
        \left\langle \hat{c}_{\sitei,\sigma}^+
            \left[\omega +(\hat{H}-E_0)-\I\eta\right]^{-1} 
         \hat{c}_{\sitei,\sigma}\right\rangle \; , \nonumber \\ 
G_{\rm UHB}(\omega ) &=& - G_{\rm LHB}(-\omega )  \label{DefGLHB} \; .
\end{eqnarray}
The last equality follows from the particle-hole symmetry, cf.~(\ref{phdef}).
Therefore, it is sufficient to
evaluate the local Green function for the lower Hubbard band
which describes the dynamics of a hole inserted into the system.

The density of states for the lower Hubbard band
can be obtained from the imaginary part of
the Green function~(\ref{DefGLHB}) for real arguments via~\cite{Fetter}
\begin{eqnarray}
D_{\rm LHB}(\omega) &=& \frac{1}{\pi} \Im G_{\rm LHB}(\omega)
\; , \nonumber
\\
&=& \frac{1}{L} \sum_{\sitei,\sigma} \left\langle 
\hat{c}_{\sitei,\sigma}^+  \delta\left(\omega+\hat{H}-E_0\right)
\hat{c}_{\sitei,\sigma} \right\rangle \; ,
\label{Dforlateruse}
\label{rangeofD}
\end{eqnarray}
with $\mu_{\rm LHB}^- \leq \omega \leq   \mu_{\rm LHB}^+ <0$.
Particle-hole symmetry results in
$D_{\rm UHB}(\omega) = D_{\rm LHB}(-\omega)$
so that the single-particle gap in the Mott--Hubbard insulator is given by
\begin{equation}
\Delta(U) = 2 | \mu_{\rm LHB}^+(U)| >0 \; .
\label{Delta}
\end{equation}

We define the (shifted) moments $M_n(U)$ of the density of states 
in the lower Hubbard band via
\begin{equation}
M_n(U)= \int_{\mu_{\rm LHB}^-}^{\mu_{\rm LHB}^+}
\diff \omega \left(\omega+\frac{U}{2}\right)^n D_{\rm LHB}(\omega) \; .
\end{equation}
In particular, from~(\ref{Dforlateruse}) we find that~\cite{Fetter}
\begin{eqnarray}
M_0(U)&=&1 \; \label{sumrule1}
,\\
M_1(U)&=&
\frac{1}{L}\left(E_0+U\frac{\partial E_0}{\partial U}\right)+\frac{U}{2} 
\label{sumrule2}
\end{eqnarray}
are two useful sum-rules which we shall employ later.

\section{Strong-coupling expansion}
\label{sec:Strong}

In this section we first summarize the Kato--Takahashi 
strong-coupling perturbation theory. Next, we prove
that the entropy density of the Mott--Hubbard insulator is finite
to all orders in perturbation theory. 
We then derive the density of states to second
order in $1/U$; for further details, see Ref.~\cite{Eva}.
Finally, we present explicit results for physical quantities
including the gap and an estimate of the critical interaction strength.

\subsection{Kato--Takahashi Perturbation Theory}
\label{subsec:Kato}

Based on Kato's degenerate perturbation theory~\cite{Kato},
Ta\-ka\-ha\-shi
developed a perturbation
expansion in $1/U$ for the Hubbard model at zero temperature~\cite{Takahashi}.
For large interaction strength~$U$, 
the set of ground states $|\psi_n\rangle$
of $\hat{H}$ in~(\ref{generalH}) can be obtained from
states $|\phi_n\rangle$ without double occupancy,
\begin{equation}
|\psi_n\rangle = \hat{\Gamma} |\phi_n\rangle \quad , \quad 
\hat{P}_0 |\phi_n\rangle =  |\phi_n\rangle \; ,
\label{gammaAction}
\end{equation}
where $\hat{P}_j$ projects onto the subspace with~$j$~double
occupancies
and $\hat{\Gamma}$ is an operator to be determined.

Eq.~(\ref{gammaAction}) is readily interpreted. In the large-coupling limit, 
$\hat{T}$ in~(\ref{generalH}) is considered to be a perturbation to~$U\hat{D}$.
Addressing the ground states we therefore start from ei\-genstates 
with zero double occupancies~$|\phi_n\rangle$ into which
the operator $\hat{\Gamma}$ 
successively introduces double occupancies
and holes to generate the ground states~$|\psi_n\rangle$ of~$\hat{H}$.
The operator $\hat{\Gamma}$ reduces all operators
to the subspace with zero double occupancies. In particular,
$\hat{\Gamma}^+\hat{\Gamma}=\hat{P}_0$ so that overlap matrix elements 
obey~\cite{Takahashi}
$\langle \psi_m | \psi_n\rangle = 
\langle \phi_m | \phi_n\rangle$.

The Schr\"odinger equation for the ground states
\begin{equation}
\hat{H}\hat{\Gamma} |\phi_n\rangle = E_0 \hat{\Gamma} |\phi_n\rangle 
\end{equation}
leads to
\begin{equation}
\hat{h}|\phi_n\rangle = E_0 |\phi_n\rangle  \quad , \quad 
\hat{h} = \hat{\Gamma}^+ \hat{H} \hat{\Gamma}\; ,
\label{deflittleh}
\end{equation}
i.e., the eigenvalue does not change under the transformation.
The creation and annihilation operators
are transformed accordingly,
\begin{equation}
\hat{c}^+_{\sitei,\sigma} \mapsto \widetilde{c}^+_{\sitei,\sigma}=
\hat{\Gamma}^+  \hat{c}^+_{\sitei,\sigma}
\hat{\Gamma} \quad , \quad 
\hat{c}_{\sitei,\sigma} \mapsto \widetilde{c}_{\sitei,\sigma}=
\hat{\Gamma}^+  \hat{c}_{\sitei,\sigma}
\hat{\Gamma} \; .
\end{equation}
The derivation of the explicit expression for~$\hat{\Gamma}$
can be found in~Refs.~\cite{Kato,Takahashi}, where it is
shown that
\begin{equation}
\hat{\Gamma} = 
\hat{P}\hat{P}_0 \left( \hat{P}_0\hat{P}\hat{P}_0 \right)^{-1/2} \; .
\label{DEFgamma}
\end{equation}
Here
\begin{eqnarray}
\hat{P} &=& \hat{P}_0 - \sum_{m=1}^{\infty} 
\sum_{{\scriptstyle r_1+\ldots+r_{m+1}=m}\atop {\scriptstyle r_i\geq 0}} 
\hat{S}^{r_1}\hat{T} \hat{S}^{r_2}
\cdots \hat{T}\hat{S}^{r_{m+1}} \; , \nonumber \\
\hat{S}^0 &=& - \hat{P}^0 \; , \\
\hat{S}^r &=& \frac{(-1)^r}{U^r}\sum_{j\neq 0} 
\frac{\hat{P}_j}{j^r} \equiv 
\frac{(-1)^r}{U^r} \widetilde{S}^r
\; .\nonumber
\end{eqnarray}
The $m$-th order term in~$\hat{P}$ contains all possible electron transfers
generated by $m$~applications of the perturbation~$\hat{T}$;
its contribution is proportional to $(1/U)^m$.
The square-root factor in~(\ref{DEFgamma}) guarantees the size-consistency
of the expansion, i.e., it eliminates the `disconnected' diagrams in
a diagrammatic formulation of the theory~\cite{Takahashi}. 
The square root of an operator is understood in terms
of its series expansion, i.e.,
\begin{equation}
\left( \hat{P}_0\hat{P}\hat{P}_0 \right)^{-1/2}
\equiv \hat{P}_0  + \sum_{m=1}^{\infty}
\frac{(2m-1)!!}{(2m)!!}
\left[  
\hat{P}_0 (\hat{P}_0 -\hat{P})\hat{P}_0 
\right]^m \; .
\end{equation}
The Green function for the lower Hubbard band becomes
\begin{equation}
G_{\rm LHB}(\omega ) = \frac{1}{L} \sum_{\sitei,\sigma}
        \left\langle \widetilde{c}_{\sitei,\sigma}^+
            \left[\omega + (\hat{h}-E_0) -\I\eta\right]^{-1} 
         \widetilde{c}_{\sitei,\sigma}\right\rangle \; ,
\label{DefGLHB2ndintermediate}
\end{equation}
where $\langle \ldots \rangle$ now implies the average over
all states~$|\phi_n\rangle$ with energy~$E_0$.

A straightforward expansion in~$1/U$,
\begin{eqnarray}
\widetilde{c}_{\sitei,\sigma} &=& \sum_{m=0}^{\infty}(-U)^{-m}
\hat{P}_0 \widetilde{c}_{\sitei,\sigma}^{(m)}\hat{P}_0 \; ,
\nonumber \\
\hat{h} &=& \frac{U}{4}(L-2\hat{N}) +
\sum_{m=0}^{\infty} U^{-m}  \hat{h}_m \; ,
\label{Hexpandone}
\end{eqnarray}
gives
\begin{eqnarray}
\hat{h}_0 &=&  \hat{P}_0\hat{T} \hat{P}_0\; ,
\label{htransformed-a} \\
\hat{h}_1 &=& -\hat{P}_0\hat{T} \widetilde{S} \hat{T}\hat{P}_0 \; ,
\label{htransformed-b}\\
\hat{h}_2 &=& \hat{P}_0\hat{T} \widetilde{S} \hat{T} \widetilde{S} \hat{T} 
\hat{P}_0 +\hat{h}_1\hat{h}_0
\label{htransformed-c}
\end{eqnarray}
for the Hamiltonian and
\begin{eqnarray}
\widetilde{c}_{\sitei,\sigma}^{(0)} &=&  \hat{c}_{\sitei,\sigma} 
\;,  \label{widetildec0}  \\
\widetilde{c}_{\sitei,\sigma}^{(1)} &=& 
\hat{c}_{\sitei,\sigma} \widetilde{S}\hat{T}
\; , \label{widetildec1}  \\
\widetilde{c}_{\sitei,\sigma}^{(2)} &=&  
\hat{c}_{\sitei,\sigma} \widetilde{S} \hat{T}\widetilde{S}\hat{T}
+ \hat{T}\widetilde{S} \hat{c}_{\sitei,\sigma} \widetilde{S}\hat{T} 
+\frac{1}{2} \hat{c}_{\sitei,\sigma} \hat{h}_1  
+ \frac{1}{2} \hat{h}_1 \hat{c}_{\sitei,\sigma}
\label{widetildec2}
\end{eqnarray}
for the annihilation operators.
In the derivation of~(\ref{htransformed-a})--(\ref{widetildec2})
we have used the fact that 
$\widetilde{c}_{\sitei,\sigma}$ in~(\ref{DefGLHB2ndintermediate}) 
acts on states $|\phi_n\rangle$ with
no holes and no double occupancies; see also Sect.~\ref{subsec:Ground}.
We have also used the fact that
in the Green function for the lower Hubbard band 
of a half-filled Bethe lattice we can express 
the Hamiltonian~$\hat{h}$ in terms of a polynomial in the
bare hole-hopping operator~$\hat{h}_0$~\cite{Eva}.

\subsection{Ground states}
\label{subsec:Ground}

To leading order in $1/U$, the states~$|\phi_n\rangle$ are eigenstates 
of the Hamiltonian $\hat{h}_1$. Because there are
no doubly occupied sites in~$|\phi_n\rangle$ and because
the system is half filled,
all sites are singly occupied (`spin-only states'). Therefore, 
$\hat{h}_1$ can be expressed in terms of spin operators.
For general hopping amplitudes $t(\sitei-\sitej)$ 
($t(\sitezero)=0$) on a regular lattice,
one finds~\cite{Takahashi,Anderson}
\begin{equation}
\hat{h}_1 = \sum_{\sitei,\sitej}2 |t(\sitei-\sitej)|^2 
\left( \hat{\vec{S}}_{\sitei} \cdot \hat{\vec{S}}_{\sitej} 
- \frac{1}{4}\right) \; .
\label{Heisenberg}
\end{equation}
In momentum space this becomes
\begin{eqnarray}
\hat{h}_1 &=& \sum_{\siteq} J(\siteq) \left[ 
\hat{\vec{S}}_{\siteq} \cdot \hat{\vec{S}}_{-\siteq} 
- \delta_{\siteq,\sitezero}\frac{L}{4}\right]
\; , \nonumber
\\
\frac{1}{2} J(\siteq) &=&  \frac{1}{L} \sum_{\sitek} \epsilon(\sitek)
\epsilon(\sitek+\siteq) \nonumber
\\
&=& \int_{-\infty}^{\infty}
 \diff \epsilon_1 \, \epsilon_1 \int_{-\infty}^{\infty} 
\diff\epsilon_2 \, \epsilon_2
\,  \rho_{\siteq}(\epsilon_1,\epsilon_2) \; .
\label{defjq} 
\end{eqnarray}
Here $\hat{\vec{S}}_{\siteq}=\sqrt{1/L} \sum_{\sitej} \exp(-\I \siteq \sitej)
\hat{\vec{S}}_{\sitej}$ is the Heisenberg spin operator in momentum space.

Since the joint density of states~(\ref{RDADq}) factorizes, 
the Hei\-senberg model~(\ref{Heisenberg}) 
in the absence of any nesting reduces to
\begin{equation}
\hat{h}_1 = J(\sitezero) \left[ \frac{\hat{\vec{S}}^2}{L}
-\frac{L}{4} \right] 
\label{2ndorderreduced}
\end{equation}
in infinite dimensions.
This shows that all global singlets are ground states with energy
$E_0^{(1)}=-J(\sitezero)L/(4U)= -\overline{\epsilon^2} L/(2U)$, 
see~(\ref{energymoments}).
For the Bethe lattice with the 
semi-elliptic density of states~(\ref{rhozero})
we have $\overline{\epsilon^2}= 1$
so that $E_0^{(1)}=-L/(2U)$. 

To first order in $1/U$, the degeneracy of the spin-only states
is not lifted because,
in the thermodynamic limit, all states with $S^z=0$ 
are also global singlets, $S=0$. 
More precisely, the ground-state entropy density is 
given by $s=\ln(2) + {\cal O}(\ln(L)/L)$~\cite{Gebhardbook}.

This degeneracy is not lifted to all orders in perturbation 
theory. We illustrate the argument in the next non-vanishing order
of the perturbation theory.
As was shown by Takahashi~\cite{Takahashi} in the presence
of particle-hole symmetry,
the Hamilton operator for the ground state to second order 
in $1/U$ vanishes, and the third-order contribution
can be cast into the form
\begin{eqnarray}
\hat{h}_3 &=& \sum_{\sitea,\siteb,\sitec} 2 |t(\sitea-\siteb)|^2
|t(\siteb-\sitec)|^2
\left( \hat{\vec{S}}_{\sitea} \cdot \hat{\vec{S}}_{\sitec} - \frac{1}{4}\right)
\nonumber \\
&&  +
\sum_{\sitea,\siteb,\sitec,\sited; \siteb\neq \sited, \sitea\neq \sitec} 
t(\sitea-\siteb)  t(\siteb-\sitec)  t(\sitec-\sited)  t(\sited-\sitea) 
\nonumber \\
&& \biggl[  
5 \left( \hat{\vec{S}}_{\siteb} \cdot \hat{\vec{S}}_{\sitec}\right)
\left( \hat{\vec{S}}_{\sitea} \cdot \hat{\vec{S}}_{\sited}\right)
+ 
5 \left( \hat{\vec{S}}_{\sitea} \cdot \hat{\vec{S}}_{\siteb}\right)
\left( \hat{\vec{S}}_{\sitec} \cdot \hat{\vec{S}}_{\sited}\right)
\nonumber \\
&& - 5 \left( \hat{\vec{S}}_{\sitea} \cdot \hat{\vec{S}}_{\sitec}\right)
\left( \hat{\vec{S}}_{\siteb} \cdot \hat{\vec{S}}_{\sited}\right)
- \frac{1}{4} \left( \hat{\vec{S}}_{\sitea} \cdot \hat{\vec{S}}_{\siteb}\right)
\nonumber \\
&& - \frac{1}{4} \left( \hat{\vec{S}}_{\siteb} \cdot 
\hat{\vec{S}}_{\sitec}\right)
- \frac{1}{4} \left( \hat{\vec{S}}_{\sitec} \cdot \hat{\vec{S}}_{\sited}\right)
- \frac{1}{4} \left( \hat{\vec{S}}_{\sited} \cdot \hat{\vec{S}}_{\sitea}\right)
\label{4thordertaka}
\\
&& 
- \frac{1}{4} \left( \hat{\vec{S}}_{\sitea} \cdot \hat{\vec{S}}_{\sitec}\right)
- \frac{1}{4} \left( \hat{\vec{S}}_{\siteb} \cdot \hat{\vec{S}}_{\sited}\right)
+ \frac{1}{16} \biggr] \nonumber
\; .
\end{eqnarray}
We have omitted terms where two sites are connected by four
hopping processes as they give a vanishing 
contribution in the limit $Z\to \infty$.
By Fourier transforming, we can ensure that the factorization
of the correlation functions as in~(\ref{RDADq}) hold.
Therefore, in infinite dimensions in the absence of perfect nesting 
eq.~(\ref{4thordertaka}) reduces to
\begin{eqnarray}
\hat{h}_3 &=& 2 [\overline{\epsilon^2}]{}^2 
\left[ \frac{\hat{\vec{S}}^2}{L} -\frac{L}{4} \right] \label{3rdorderreduced}
 \\
&& + \left(\overline{\epsilon^4} -2 [\overline{\epsilon^2}]{}^2 \right)
\left[  \frac{5}{L} \left( \frac{\hat{\vec{S}}^2}{L}\right)^2 
- \frac{3}{2} \frac{\hat{\vec{S}}^2}{L}
+\frac{L}{16} \right] \, .
\nonumber
\end{eqnarray}
The second term describes the (usually ferromagnetic)
ring exchange whereas the first term favors ground states with spin zero.

Like $\hat{h}_1$ in~(\ref{2ndorderreduced}), 
$\hat{h}_3$ in~(\ref{3rdorderreduced}) is solely a function 
of the operator for the total spin.
This argument is readily generalized to higher orders in
the perturbation expansion, i.e., $\hat{h}$
is a function of~$\hat{\vec{S}}^2$ where the leading
term favors total spin zero.
Therefore, as long as the perturbation series in $1/U$ converges,
the degeneracy of the ground state is not lifted.
As a consequence of its huge degeneracy, $s=\ln(2)$,
each lattice site is equally likely to be occupied by an electron with
spin $\uparrow$ or $\downarrow$, 
irrespective of the spin on any other lattice site.
This is a characteristic feature of the Mott--Hubbard insulator
in the limit of high dimensions.

On the Bethe lattice there are no closed loops, and
the ring exchange is absent, $\overline{\epsilon^4} =
2 \overline{\epsilon^2}=2$. Thus, the ground-state energy to third
order becomes $E_0^{(3)}=-L/(2 U^3)$.

\subsection{Green function to second order}
\label{subsec:Greencalc}

As shown in Ref.~\cite{Eva}, to second order in $1/U$
the Green function for the
lower Hubbard band can be expressed in the form
\begin{equation}
G_{\rm LHB}(\omega) = \frac{1}{L} \sum_{\sitei,\sigma}   \left\langle 
\hat{c}_{\sitei,\sigma}^+ 
s(\hat{h}_0,U)
 \left[z + g(\hat{h}_0,U)\right]^{-1}
\hat{c}_{\sitei,\sigma} 
\right\rangle \;,
\label{GofzDef}
\end{equation}
where $z=\omega+U/2-\I \eta$, and $s(\epsilon,U)$ and $g(\epsilon,U)$ 
are polynomials in $\epsilon$ and $1/U$.
They are called
the shape-correction factor and the gap-renormalization
factor, respectively.
Up to second order in $1/U$, the general structure of~$\hat{h}$
in~(\ref{htransformed-a})--(\ref{htransformed-c}) 
and of $\widetilde{c}_{\sitei,\sigma}$
in~(\ref{widetildec0})--(\ref{widetildec2}) implies
\begin{eqnarray}
s(\epsilon,U) &=& 1+\frac{s_{1,1}\epsilon}{U}
+\frac{s_{2,2}\epsilon^2+s_{2,0}}{U^2}
\nonumber \; , \\
g(\epsilon,U) &=&\epsilon + \frac{g_{1,2}\epsilon^2+g_{1,0}}{U}+
\frac{g_{2,3}\epsilon^3+g_{2,1}\epsilon}{U^2} 
\end{eqnarray}
because each electron transfer operator~$\hat{T}$ 
generates one bare hole-hopping
operator~$\hat{h}_0$ at most. 

\subsubsection{Gap-renormalization factor}
\label{subsubsec:Gap}

We determine the coefficients $g_{i,j}$ first. 
To this end, we set
$s(\epsilon,U) \equiv 1$ in~(\ref{GofzDef}) and
expand this equation in $1/U$. 
We equate
the result of this expansion with the corresponding expression
from~(\ref{DefGLHB2ndintermediate}). To first order in $1/U$ we find
\begin{eqnarray}
\lefteqn{\frac{1}{L} \sum_{\sitei,\sigma} 
\left\langle \hat{c}_{\sitei,\sigma}^+
            \left[z+\hat{h}_0\right]^{-2}
\left[ g_{1,2}(\hat{h}_0)^2+g_{1,0}\right]
          \hat{c}_{\sitei,\sigma}\right\rangle =
} 
\\
&& 
\frac{1}{L} \sum_{\sitei,\sigma} 
\left\langle \hat{c}_{\sitei,\sigma}^+
            \left[z+\hat{h}_0\right]^{-1}
\left(\hat{h}_1+\frac{L}{2U}\right)
            \left[z+\hat{h}_0\right]^{-1}
          \hat{c}_{\sitei,\sigma}\right\rangle \; .
\nonumber 
\end{eqnarray}
As we need to fix two coefficients only, we further expand this expression
in $1/z$. Keeping the first two non-trivial orders we arrive at
\begin{eqnarray}
g_{1,2}+g_{1,0} &=& 
\frac{1}{L} \sum_{\sitei,\sigma} 
\left\langle \hat{c}_{\sitei,\sigma}^+
\left[ g_{1,2}(\hat{h}_0)^2+g_{1,0}\right]
          \hat{c}_{\sitei,\sigma}\right\rangle 
\nonumber \\
&=& 
\frac{1}{L} \sum_{\sitei,\sigma} 
\left\langle \hat{c}_{\sitei,\sigma}^+
\left(\hat{h}_1+\frac{L}{2}\right)
          \hat{c}_{\sitei,\sigma}\right\rangle 
 \; ,
\label{h1first}
\end{eqnarray}
and 
\begin{eqnarray}
2g_{1,2}+g_{1,0} &=&
\frac{1}{L} \sum_{\sitei,\sigma} 
\left\langle \hat{c}_{\sitei,\sigma}^+
\left[ g_{1,2}(\hat{h}_0)^4+g_{1,0}(\hat{h}_0)^2\right]
          \hat{c}_{\sitei,\sigma}\right\rangle 
\nonumber \\
&=&
\frac{1}{L} \sum_{\sitei,\sigma} 
\left\langle \hat{c}_{\sitei,\sigma}^+
\left(\hat{h}_1+\frac{L}{2}\right)
 (\hat{h}_0)^2
          \hat{c}_{\sitei,\sigma}\right\rangle 
 \; .
\label{h1second}
\end{eqnarray}
It is not difficult to calculate the expectation values in
these equations from the definition~(\ref{htransformed-b}).
Recall that $\langle \ldots \rangle$ implies the average over all spin-only
states. In eq.~(\ref{h1first}) we note that an intermediate nearest-neighbor
pair of hole and double occupancy can be created with
probability $1/2$ by $\hat{h}_1$
everywhere but at site~$\sitei$.
In the sum over the whole lattice,
two contributions are therefore missing.
Altogether, $\hat{h}_1$ generates the contribution
$-(L-2)/2$ in~(\ref{h1first}), which, together with the first-order 
contribution
$L/2$ from the ground-state energy, results in $g_{1,2}+g_{1,0}=1$.
In~(\ref{h1second}) the three-site term in $\hat{h}_1$ contributes
with probability~1/2. Thus,
$2g_{1,2}+g_{1,0}=1-1/2=1/2$. Therefore, $g_{1,2}=-1/2$ and $g_{1,0}=3/2$,
and $\hat{h}_1^{\rm eff}\equiv g_{1,2}\hat{h}_0^2+ g_{1,0}=
-(\hat{h}_0^2-3)/2$.

In second order, we split the contributions arising from $\hat{h}_2$
and those from the correction terms. Let $\hat{C}=1/(z+\hat{h}_0)$
and $\hat{h}_1'=\hat{h}_1 -E_0^{(1)} -\hat{h}_1^{\rm eff}$, 
$\hat{h}_2^{\rm eff}\equiv g_{2,3}\hat{h}_0^3+g_{2,1}\hat{h}_0
=\hat{h}_{2a}^{\rm eff}+\hat{h}_{2b}^{\rm eff}$, then
\begin{eqnarray}
\frac{1}{L} \sum_{\sitei,\sigma} 
\left\langle \hat{c}_{\sitei,\sigma}^+ 
\hat{h}_{2a}^{\rm eff} \hat{C}^2
\hat{c}_{\sitei,\sigma}\right\rangle 
&=&
\frac{1}{L} \sum_{\sitei,\sigma} 
\left\langle \hat{c}_{\sitei,\sigma}^+
\hat{C} 
\hat{h}_2
\hat{C}
\hat{c}_{\sitei,\sigma}\right\rangle  \; ,\\
\frac{1}{L} \sum_{\sitei,\sigma} 
\left\langle \hat{c}_{\sitei,\sigma}^+ 
\hat{h}_{2b}^{\rm eff}\hat{C}^2
\hat{c}_{\sitei,\sigma}\right\rangle 
&= &
- \frac{1}{L} \sum_{\sitei,\sigma} 
\left\langle \hat{c}_{\sitei,\sigma}^+
\hat{C}
\hat{h}_1'
\hat{C}
\hat{h}_1'
\hat{C}
\hat{c}_{\sitei,\sigma}\right\rangle  \; .\nonumber \\
&& 
\label{twospinflipterms}
\end{eqnarray}
Expanding in $1/z$, the corresponding equations 
for $\hat{h}_{2a}^{\rm eff}$ read
\begin{eqnarray}
2g_{2,3;a}+g_{2,1;a} 
&=& 
\frac{1}{L} \sum_{\sitei,\sigma} 
\left\langle \hat{c}_{\sitei,\sigma}^+
\hat{h}_2\hat{h}_0
          \hat{c}_{\sitei,\sigma}\right\rangle 
\; ,
\\
g_{2,3;a} 
&=& 
\frac{1}{L} \sum_{\sitei,\sigma} 
\left\langle \hat{c}_{\sitei,\sigma}^+
\hat{h}_2 [(\hat{h}_0)^3-2\hat{h}_0]
          \hat{c}_{\sitei,\sigma}\right\rangle 
\, . \, 
\label{h2second}
\end{eqnarray}
Note that in~(\ref{h2second}) the operator $(\hat{h}_0)^2-1$ transfers
the hole at~$\sitei$ into its second neighbor shell.
After some calculations we find $2g_{2,3;a}+g_{2,1;a}=-3/2$ and
$g_{2,3;a}=3/4$ so that $g_{2,1;a}=-3$.

The correction terms were correctly taken into account
for the Falicov--Kimball model in Ref.~\cite{Eva}, but were
erroneously omitted for the Hubbard model.
The operator $\hat{h}_1'$
describes the motion of a hole by two sites whereby a spin-flip
is induced. In~(\ref{twospinflipterms}) the hole hops from
site~$\sitei$ to some site~$\sitej$ where the hole
induces a spin-flip. This needs to be healed after the hole has made
an excursion into the lattice so that
\begin{eqnarray}
\frac{1}{L} \sum_{\sitei,\sigma} 
\left\langle \hat{c}_{\sitei,\sigma}^+ 
\hat{h}_{2b}^{\rm eff}\hat{C}^2
\hat{c}_{\sitei,\sigma}\right\rangle 
&=& -G_0(z) 
\frac{1}{L} \!\sum_{\sitei,\sigma} 
\left\langle \hat{c}_{\sitei,\sigma}^+ 
\hat{C}
(\hat{h}_1')^2
\hat{C}
\hat{c}_{\sitei,\sigma}\right\rangle 
\nonumber \\
&=& -\frac{3}{4} G_0(z) 
\frac{1}{L} \sum_{\sitei,\sigma} 
\left\langle \hat{c}_{\sitei,\sigma}^+ 
\hat{C}^2
\hat{c}_{\sitei,\sigma}\right\rangle 
\nonumber
\\
&=& \frac{3}{8}
\frac{1}{L} \sum_{\sitei,\sigma} 
\left\langle \hat{c}_{\sitei,\sigma}^+ 
\hat{h}_0 \hat{C}^2
\hat{c}_{\sitei,\sigma}\right\rangle  \; .
\end{eqnarray}
This leads to $\hat{h}_{2b}^{\rm eff}=(3/8)\hat{h}_0$,
i.e., $g_{2,3;b}=0$, $g_{2,1;b}=3/8$.

Altogether, $g_{2,3}=g_{2,3;a}+g_{2,3;b}=3/4$ and
$g_{2,1}=g_{2,1;a}+g_{2,1;b}=-21/8$, and
$\hat{h}_2^{\rm eff}=3\hat{h}_0[2 (\hat{h}_0)^2-7]/8$.
The gap-renormalization factor becomes
\begin{equation}
g(\epsilon,U)=\epsilon 
-\frac{\epsilon^2-3}{2U}
+\frac{3 \epsilon \left(2 \epsilon^2-7\right)}{8 U^2}
\label{finalG}
\end{equation}
through second order in $1/U$.

\subsubsection{Shape-correction factor}
\label{subsubsec:Shape}

In order to determine the coefficients $s_{i,j}$
we first employ the sum rules~(\ref{sumrule1}) and~(\ref{sumrule2}).
We use $g(\epsilon,U)$ from the previous subsection and find
through terms of order $1/U^3$,
\begin{eqnarray}
1 &= & \int \diff\epsilon  \rho_0(\epsilon) s(\epsilon,U) \; ,
\label{MOuse}\\
\frac{1}{U^3} &=& 
- \int \diff \epsilon \rho_0(\epsilon) s(\epsilon,U) 
g(\epsilon,U)\; ,
\label{M1use}
\end{eqnarray}
because $E_0/L=-U/4 - 1/(2U) -1/(2U^3) +{\cal O}(U^{-5})$.
This immediately gives 
$s_{2,2}=-s_{2,0}$ from~(\ref{MOuse}). From~(\ref{M1use})
we obtain $s_{1,1}=-1$.

For the other coefficients we expand~(\ref{GofzDef})
and~(\ref{DefGLHB2ndintermediate}) in $1/U$ and in $1/z$. 
The shape-correction coefficients to second order follow from
the equation
\begin{eqnarray}
s_{2,2} &=& 2 
\frac{1}{L} \sum_{\sitei,\sigma} 
\left\langle \hat{c}_{\sitei,\sigma}^+
\left[ (\hat{h}_0)^2-1\right]
          \widetilde{c}_{\sitei,\sigma}^{(2)}\right\rangle 
\nonumber \\
&& 
+ \frac{1}{L} \sum_{\sitei,\sigma} 
\left\langle \left(\widetilde{c}_{\sitei,\sigma}^{(1)}\right)^+
\left[ (\hat{h}_0)^2-1\right]
          \widetilde{c}_{\sitei,\sigma}^{(1)}\right\rangle 
\\
&=& 2 \left[ \frac{3}{4}+\frac{1}{2} -\frac{1}{4}\right] + \frac{1}{4} 
= \frac{9}{4} \; .
\nonumber
\end{eqnarray}
We hereby correct a minor mistake in Ref.~\cite{Eva} where $s_{2,2}=5/4$
was reported.

Altogether the shape-correction factor becomes
\begin{equation}
s(\epsilon,U)=1 -\frac{\epsilon}{U}
+\frac{9\left(\epsilon^2-1\right)}{4 U^2}
\label{finalS}
\end{equation}
up to and including second order in $1/U$.

\subsection{Physical quantities}
\label{subsec:physquant}

\subsubsection{Density of states}
\label{subsubsec:dos3}

The density of states to order $(1/U)^0$ 
is readily obtained. It results from the motion of a single hole
created at site~$\sitei$ which moves through the Bethe lattice via
$\hat{h}_0$ and returns to~$\sitei$. The hole has not disturbed
the spin background after its return to~$\sitei$, i.e., 
its motion appears to be free. Therefore, 
$\rho_h(\omega)=\rho_0(\omega+U/2)$~\cite{Eva,MVSchmit}.
Now that we have expressed $G(\omega)$ in~(\ref{GofzDef})
in terms of $\hat{h}_0$ we can easily read off the expression 
for the density of states,
\begin{equation}
D_{\rm LHB}(\omega) = \int_{-2}^{2} \diff \epsilon \rho_0(\epsilon)
s(\epsilon,U) \delta\left(\omega+U/2 +g(\epsilon,U)\right) \,.
\label{DOSfinalres}
\end{equation}
The second-order expressions for the gap-renormalization factor~$g(\epsilon,U)$
and the shape-correction factor~$s(\epsilon,U)$ are given in~(\ref{finalG})
and~(\ref{finalS}).

To all orders in the $1/U$ expansion, 
the density of states increases algebraically near the gap,
\begin{eqnarray}
D_{\rm UHB}(\omega) &\sim& \left(\omega-\frac{\Delta(U)}{2}
\right)^{\alpha_{\rm sc}} 
\quad , \quad \omega \to \frac{\Delta(U)}{2}
\label{dosexponenta}
\; ,\\
\alpha_{\rm sc}&=&\frac{1}{2}
\; . 
\label{dosexponentb}
\end{eqnarray} 
This reflects the fact that $\rho_0(\epsilon)$ displays
a square-root increase at the band edges. As long as the shape-correction
factor $s(\epsilon,U)$ remains finite for $|\epsilon|\to 2$, 
the shape of the Hubbard bands near the edges remains the same for all
$U>U_{\rm c}$. This behavior is qualitatively and quantitatively
the same as in
the exactly solvable Falicov--Kimball model~\cite{PvD}.

\subsubsection{Single-particle gap and width of the Hubbard bands}
\label{subsubsec:gap3}

As seen from~(\ref{DOSfinalres})
the density of states becomes zero at 
\begin{equation}
\mu_{\rm LHB}^{\pm}(U) + \frac{U}{2} + g(\mp 2, U) =0\; .
\end{equation}
Using~(\ref{Delta}) we find for the gap
\begin{equation}
\Delta(U) = U +2 g(-2,U) = U-4 -\frac{1}{U} -\frac{3}{2 U^2}
+{\cal O}\left(U^{-3}\right) \; .
\label{gapfinal3rd}
\label{gapfinal2nd}
\end{equation}
The Hubbard bands have the width 
\begin{equation}
W_{\rm LHB}=W_{\rm UHB}= 
\mu_{\rm LHB}^{+}-\mu_{\rm LHB}^{-} = 4+\frac{3}{2 U^2}  
\end{equation}
up to second order in $1/U$. The $1/U$ expansion therefore provides
a very good guess for the support for the density of states
as a function of frequency
in the Mott--Hubbard insulator for all $U$. We shall profit
from this result in our numerical calculations.

The gap appears to converge rapidly as a function of $1/U$.
Using the expansion parameter $x=2/U$, 
we may write~(\ref{gapfinal3rd}) as
\begin{equation}
\Delta(U) = U-4 - \frac{1}{2} \left[ x + \frac{3}{4} x^2 + \lambda
x^3 +{\cal O}(x^4)\right] \; .
\end{equation}
Preliminary results from our calculations to third order
indicate that indeed $\lambda\approx 1$.

The critical interaction strength is determined from
$\Delta(U)=0$. If we denote by $U_{\rm c}^{(m)}$ the critical
interaction strength at which the $m$-th order expression for $\Delta(U)$
vanishes, we find using $\lambda=1$
\begin{eqnarray}
U_{\rm c}^{(0)} &=& 4 \; , \nonumber \\
U_{\rm c}^{(1)} &=& 4.236 \, [5.9\%]\; ,\nonumber \\
U_{\rm c}^{(2)} &=& 4.313 \, [1.8\%] \;, \nonumber \\ 
U_{\rm c}^{(3)} &\approx& 4.357 \, [\approx 1.0\%] \; .
\end{eqnarray}
The numbers in brackets give the percentage change to the result
from the previous order. 
Under the assumptions that all higher-order contributions 
go in the same direction, and decay similarly fast, we \emph{conjecture\/}
that 
\begin{equation}
U_{\rm c}^{\rm sc} = 4.40 \pm 0.09 \label{Ucguess} \; ,
\end{equation}
which allows for another 3\% increase of $U_{\rm c}$
with respect to the third-order result.
We shall see in Sects.~\ref{sec:Strong} and~\ref{sec:FEED}
that this estimate is in excellent agreement
with our results from numerical calculations.

This analysis may also be carried out for the
exactly solvable Falicov--Kimball model~\cite{PvD}, where
\begin{eqnarray}
U_{\rm c,FK}^{(0)} &=& 2.828 \; , \nonumber \\
U_{\rm c,FK}^{(1)} &=& 2.414 \, [-14.6\%]\; , \nonumber \\
U_{\rm c,FK}^{(2)} &=& 2.242 \, [-7.15\%] \; , \nonumber \\
U_{\rm c,FK}^{(3)} &=& 2.167 \, [-3.31\%] \; , \nonumber \\
U_{\rm c,FK} &=& 2 \, [-8.35\%] \; . 
\end{eqnarray}
It appears to us that for the Falicov--Kimball model the
convergence of the critical interaction strength
is not quite as fast as for the Hubbard model. The relative changes
are about a factor of three smaller for the Hubbard case, and
correspondingly, we have reason to expect that for the Hubbard model
the exact result is within 3\% of the third-order result. 

\subsubsection{Self-energy and momentum distribution}
\label{subsubsec:selfenergy}

Up to an overall coefficient, the single-particle density of states 
is identical to the imaginary part
of the Green function~(\ref{rangeofD}). The Kramers--Kronig
relation provides the real part as
\begin{equation}
\Re G(\omega)
= {\rm P} \!\int_{\mu_{\rm LHB}^-}^{\mu_{\rm LHB}^+}
\diff \omega'
D_{\rm LHB}(\omega') \left( \frac{1}{\omega-\omega'}
+ \frac{1}{\omega+\omega'}\right) \; .
\end{equation}
The local Green function per spin
$G_{\sigma}(\omega)=G(\omega)/2$ and the single-particle 
self-energy $\Sigma_{\sigma} (\omega) = -\Sigma_{\sigma}(-\omega)$
are related by
\begin{eqnarray}
G_{\sigma}(\omega) 
&=& 
\int_{-\infty}^{\infty} \diff\epsilon \rho_0(\epsilon)
\frac{1}{\omega-\epsilon-\Sigma_{\sigma}(\omega)+\I \eta \, \sgn (\omega)}
\; \nonumber ,\\
&=& G_{\sigma}^{0}(\omega-\Sigma_{\sigma} (\omega))
\end{eqnarray}
in infinite dimensions, and 
\begin{equation}
G_{\sigma}^{0}(z)=\frac{z}{2}  \left[  
1- \sqrt{1-\frac{4}{z^2}}
\, \right]
\label{SigmafromG}
\end{equation}
holds for the Bethe lattice~\cite{Economou} so that
\begin{equation}
\omega-\Sigma_{\sigma} (\omega) = G_{\sigma}(\omega) + 
\frac{1}{G_{\sigma}(\omega)} \; .
\end{equation}
The real and imaginary part of the self-energy
are then readily obtained as 
\begin{eqnarray}
\Re\Sigma_{\sigma} (\omega)&=&\omega - \Re G_{\sigma}(\omega) 
\left[1+ \frac{1}{(\Re G_{\sigma}(\omega) )^2
+ (\Im G_{\sigma}(\omega) )^2} \right] \, ,
\nonumber \\
&& \\
\Im\Sigma_{\sigma} (\omega)&=& -\Im G_{\sigma}(\omega) 
\left[1- \frac{1}{(\Re G_{\sigma}(\omega) )^2
+ (\Im G_{\sigma}(\omega) )^2} \right]
 \; . \nonumber\\
&&  
\end{eqnarray}
Note that this self-energy is not perturbatively linked to the 
weak-coupling limit.
For example, its imaginary part displays a peak $\delta(\omega)$ and its
real part diverges proportional to $(-1/\omega)$ as $\omega\to 0$.

The single-particle spectral function $A_{\sigma} (\epsilon;\omega)$
is defined by
\begin{eqnarray}
A_{\sigma} (\epsilon;\omega) &=& -\frac{1}{\pi} \sgn(\omega) 
\Im\left( \frac{1}{\omega-\epsilon-\Sigma_{\sigma}(\omega)}
\right) 
\label{Defspectral}
\\
&=& -\frac{1}{\pi} \sgn(\omega) 
\frac{\Im \Sigma_{\sigma}(\omega)}{(\omega-\epsilon-
\Re \Sigma_{\sigma}(\omega))^2+
(\Im \Sigma_{\sigma}(\omega))^2} \, .
\nonumber 
\end{eqnarray}
It does not display a quasi-particle contribution. The momentum distribution
\begin{equation}
n_{\sigma}(\epsilon) = \int_{\mu_{\rm LHB}^-}^{\mu_{\rm LHB}^+}
\diff \omega A_{\sigma} (\epsilon;\omega) 
\label{defnofepsilon}
\end{equation}
depends on momentum implicitly via $\epsilon\equiv\epsilon(\sitek)$.
In the insulating phase, the momentum distribution is an analytic function
of $\epsilon$. Since $n_{\sigma}(\epsilon)=1-n_{\sigma}(-\epsilon)$
due to particle-hole symmetry, the Taylor expansion around $\epsilon=0$
has the form
\begin{equation}
n(\epsilon) = \frac{1}{2} - c_1 \epsilon -c_3 \epsilon^3 \ldots
\; .
\label{nofkcoeff}
\end{equation}
For the Bethe lattice we obtain $c_1 = 1/(2U)+ 
{\cal O}(U^{-3})$~\cite{Takahashi}, and, in general,
$c_{2n-1}={\cal O}(U^{1-2n})$.

\section{Exact diagonalization with fixed energy interval (FE-ED)}
\label{sec:FEED}

In this section, we first discuss the single-impurity model
onto which the Hubbard model can be mapped in the limit
of infinite dimensions. Then we propose the Fixed-Energy 
Exact Diagonalization (FE-ED) as a new scheme 
for the numerical solution of the DMFT equations.
The results of this approach corroborate our findings
from Sect.~\ref{sec:Strong}.

\subsection{Dynamical Mean-Field Theory (DMFT)}
\label{subsec:DMFT}

In the limit of infinite dimensions~\cite{MV} and under the conditions
of translational invariance and convergence of perturbation theory in strong
and weak coupling, lattice models for correlated electrons 
can be mapped onto single-impurity models
which need to be solved self-con\-sist\-ently
\cite{BrandtMielsch,Jarrell,RMP}. Unfortunately,
these impurity models cannot be solved analytically.

For an approximate numerical treatment, various different implementations
are conceivable; see, e.g., Ref.~\cite{Potthoff} for a recent implementation.
One realization is the single-impurity Anderson model in `star geometry',
\begin{eqnarray}
\hat{H}_{\rm SIAM} &=&\sum_{\ell=1;\sigma}^{n_s-1}
\epsilon_{\ell} \hat{\psi}_{\sigma;\ell}^+\hat{\psi}_{\sigma;\ell}
+ U \left( \hat{d}_{\uparrow}^+\hat{d}_{\uparrow} -\frac{1}{2}\right)
\left( \hat{d}_{\downarrow}^+\hat{d}_{\downarrow} -\frac{1}{2}\right) 
\nonumber \\
&& + \sum_{\sigma} \sum_{\ell=1}^{n_s-1} V_{\ell}
\left( \hat{\psi}_{\sigma;\ell}^+\hat{d}_{\sigma} + 
\hat{d}_{\sigma}^+\hat{\psi}_{\sigma;\ell}\right) \; ,
\label{SIAMns}
\end{eqnarray}
where $V_{\ell}$ are real, positive hybridization matrix elements.
The model describes the hybridization of an impurity site 
with Hubbard interaction to $n_s-1$ bath sites without interaction 
at energies $\epsilon_{\ell}$ with
$0<\epsilon_1<\epsilon_2 <\ldots < \epsilon_{(n_s-1)/2}$. 
In order to ensure particle-hole
symmetry, we have to set $\epsilon_{\ell}=-\epsilon_{n_s-\ell}$
and $V_{\ell}=V_{n_s-\ell}$
for $\ell=(n_s+1)/2,\ldots,n_s-1$. Moreover, since we are interested 
in the Mott--Hubbard 
insulator, we only use 
odd $n_s$ so that there is no bath state at $\epsilon=0$.

For a given set of $(n_s-1)$ parameters $(\epsilon_{\ell}, V_{\ell})$
the model~(\ref{SIAMns}) defines a many-body problem for which the 
single-particle Green function 
\begin{equation}
G_{\sigma}^{(n_s)}(t) = -\I \left\langle \hat{{\cal T}} \left[
\hat{d}_{\sigma}(t) \hat{d}_{\sigma}^+\right]
\right\rangle_{\rm SIAM} 
\label{GSIAMfinite}
\end{equation}
can be calculated numerically for $n_s \leq 15$ with the (dynamical) 
Lanczos technique. In~(\ref{GSIAMfinite}) $\langle \ldots\rangle_{\rm SIAM}$ 
implies the ground-state expectation value within the single-impurity model.
Typically, the imaginary part of the Green function displays
$n_s$ peaks with large weight and many other small peaks
whose number depends on the number of states $n_{\rm L}$ kept in the Lanczos
diagonalization.

Ultimately, we are interested in the limit $n_s \to\infty$ where the 
hybridization function
\begin{equation}
H^{(n_s)}(\omega) 
= \sum_{\ell=1}^{n_s-1} 
\frac{V_{\ell}^2}{\omega-\epsilon_{\ell}+\I \eta\, \sgn(\omega)} 
\end{equation}
should smoothly approach 
the hybridization function of the \emph{continuous} problem, 
\begin{equation}
H(\omega) = \lim_{n_s\to\infty}H^{(n_s)}(\omega) \; .
\end{equation}
Correspondingly, the Green function should fulfill
\begin{equation}
G_{\sigma}(\omega)
= \lim_{n_s\to\infty}
G_{\sigma}^{(n_s)}(\omega)
 \; .
\end{equation}
At self-consistency, the Green function of the impurity problem
describes the Hubbard model in infinite dimensions.
As shown in~\cite{RMP},
on the Bethe lattice the hybridization function must obey 
the simple relation
\begin{equation}
H(\omega) = G_{\sigma}(\omega)
\label{selfcons}
 \; .
\end{equation}
This equation closes the self-consistency cycle for the continuous problem:
we have to choose bath energies and hybridizations in such a way that the
single-particle Green function and 
the hybridization function fulfill~(\ref{selfcons}).

\subsection{Implementation of the FE-ED}
\label{subsec:FEEDimplementation}

Any numerical scheme for the DMFT faces two problems. First, and foremost,
the DMFT requires the Green function $G_{\sigma}(\omega)$ on
the whole, continuous frequency axis but only the information for
$(n_s-1)$ energies $\epsilon_{\ell}$ is provided.
This is in contrast to standard numerical problems in many-body theory where
a given energy interval, typically of the order of the band-width~$W$,
is resolved to accuracy $W/n_s$. Concomitantly, in numerical DMFT calculations
it is not a priori clear how observables scale as a function of $1/n_s$.
Moreover, there can be more than one self-consistent set of parameters 
$(\epsilon_{\ell}, V_{\ell})$ for fixed $n_s$.

Second, the self-consistency condition~(\ref{selfcons}) holds 
for the continuous $H(\omega)$ and $G_{\sigma}(\omega)$, 
not for their discretized
counterparts. Therefore, the various ED methods will differ in the
way the information is extracted from 
$G_{\sigma}^{(n_s)}(\omega)$ 
in one iteration in order to specify the input parameters 
$(\epsilon_{\ell}, V_{\ell})$ for the next iteration.
This is another source of ambiguity. Since the scaling as a function of
$1/n_s$ is not clear, it cannot be guaranteed that different schemes
will ultimately coincide as $n_s \to\infty$.

In this work, we use the results of the $1/U$ expansion to circumvent
the first problem. We have seen in Sect.~\ref{sec:Strong} that the 
Hubbard bands are distributed symmetrically around $\omega =0$ with width
$W_{\rm LHB}=W_{\rm UHB}=W^*\approx W$. Thus, we actually need to resolve 
a \emph{finite} frequency interval. 
In practice, we use $W^*=4.5$ as our maximum band-width.
Then, we determine 
the onset of the upper Hubbard band self-consistently, 
$\Delta(U)/2$; see below.
To this end, we start with some input guess $\Delta_{\rm in}(U)$.
In practice, we use the second-order estimate of the gap
from~(\ref{gapfinal2nd}) to speed up the convergence
of this outer self-consistency cycle.

We choose to discretize the Hubbard bands equidistantly,
i.e., we fix the energies $\epsilon_{\ell}$ for $1\leq \ell \leq (n_s-1)/2$ by
\begin{equation}
\epsilon_{\ell}= \frac{\Delta_{\rm in}(U)}{2} 
+ \left(\ell-\frac{1}{2}\right)
\delta W \quad ; \quad \delta W= \frac{2W^*}{n_s-1} \; .
\end{equation}
For $n_s \leq 15$ and $U>W$ 
none of the $\epsilon_{\ell}$ can move outside the Hubbard bands. 
By fixing the energies at the centers of the equidistant intervals 
\begin{equation}
I_{\ell}=
[\frac{\Delta_{\rm in}(U)}{2} + (\ell-1) \delta W, 
\frac{\Delta_{\rm in}(U)}{2} + \ell \delta W] \; , 
\end{equation}
$1\leq \ell \leq (n_s-1)/2$, we can be sure that our
resolution of the Hubbard bands becomes increasingly better as $n_s$ increases.

We address the second problem in the following way. 
We apply a constant broadening 
of width $\delta W$ to the individual peaks 
in $\Im G_{\sigma}^{(n_s)}(\omega_r)$, 
$r=1,\ldots,n_{\rm L}$.
Then, we collect the weight 
into the intervals $I_{\ell}$ and assign this weight
to $w_{\ell}=V_{\ell}^2$. 
In the Caffarel--Krauth scheme~\cite{CK,Ono} a fictitious temperature and
a $\chi^2$-fitting procedure is used. We have found that our simple
approach is equally suitable.

Typically, the weight of the peaks at energies
outside the Hubbard bands is very small. Thus we set 
\begin{eqnarray}
w_{\ell}
&=& \int_{I_{\ell}} \diff \omega 
\sum_{r=1}^{n_{\rm L}} 
\Im G_{\sigma}^{(n_s)}(\omega_r) \nonumber
\\
&& \times \frac{\Theta(\omega - \omega_r+ \delta W/2)
- \Theta(\omega - \omega_r- \delta W/2)}{\delta W}
\; .
\label{recollectionofweight}
\end{eqnarray}
In order to generate an educated input guess for $V_{\ell}=\sqrt{w_{\ell}}$
we use the results from second-order perturbation theory
in~(\ref{recollectionofweight}).

Now that we have determined the input energies~$\epsilon_{\ell}$ and
hybridizations $V_{\ell}$, we can start the inner self-consistency cycle.
We perform a dynamical
Lanczos diagonalization which gives $\Im G_{\sigma}^{(n_s)}(\omega_r) $
for $n_{\rm L}$ Lanczos frequencies. Most of the peaks carry
little weight, only $n_s$ of them are significant.
Using~(\ref{recollectionofweight}), the Green function provides
the hybridization elements for the next iteration
of the inner self-consistency cycle.
We have checked that, for fixed $n_s$, 
a unique solution for $w_l$ is found for various starting choices. 

\begin{figure}[ht]
\resizebox{\columnwidth}{!}{%
\includegraphics{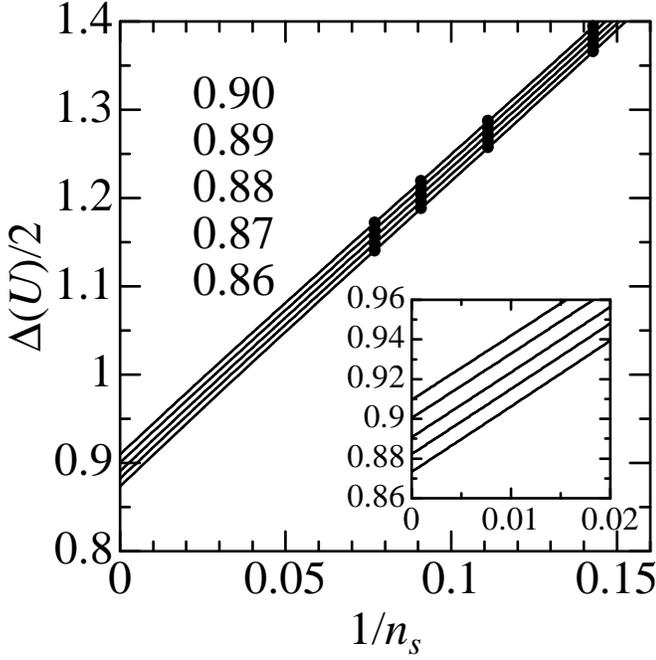}
}
\caption{Gap $\Delta^{E;(n_s)}(U)$ from~(\protect\ref{FE-EDgap}) for $U=6$
as a function of $1/n_s$ for various $\Delta_{\rm in}$.}
\label{Fig:FE-EDscheme-energy}
\end{figure}

After we have determined the parameters $V_{\ell}$
self-con\-sist\-ently for fixed $n_s$ and a given~$U$, 
we may calculate the single-particle
gap from the difference in
the ground-state energy at half band-filling, $E_0^{(n_s)}(N;U)$,
and the ground-state energy for one particle more than
half band-filling, $E_0^{(n_s)}(N+1;U)$, as
\begin{equation}
\Delta^{E;(n_s)}(U) 
= 2 \left( E_0^{(n_s)}(N+1;U)
- E_0^{(n_s)}(N;U)\right)
\label{FE-EDgap}
\; .
\end{equation}
After the extrapolation to the thermodynamic limit, 
\begin{equation}
\Delta^{E}(U) = \lim_{n_s \to\infty}\Delta^{E;(n_s)}(U)
\; , 
\label{deltagsns}
\label{energycriterion}
\end{equation}
we obtain a new estimate for the onset of the Hubbard bands,
and we could restart the outer self-consistency cycle.
Unfortunately, the extrapolation $n_s\to\infty$ has limited accuracy,
and it is difficult to obtain a truly converged solution which fulfills
$\Delta^{E}(U) = \Delta_{\rm in}(U)$ (`energy criterion'), 
even at very large values, e.g., $U=14$. 
An example for $U=6$ is shown in Fig.~\ref{Fig:FE-EDscheme-energy}.
It is seen that the results around $\Delta_{\rm in}=0.88$
are almost but not quite converged. 

\begin{figure}[ht]
\resizebox{\columnwidth}{!}{%
\includegraphics{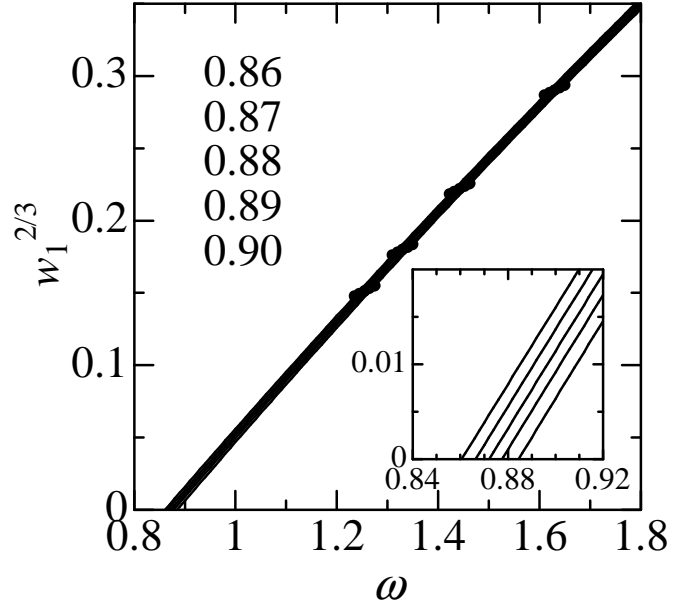}
}
\caption{Weight of the first peak in the upper Hubbard band 
for $n_s=7,9,11,13$, $U=6$, and various $\Delta_{\rm in}$.
For a square-root increase of the
density of state, $w_1^{2/3}$ should scale linearly with $1/n_s$,
see~(\protect\ref{weightcriterion}). The intersection of the
extrapolated curves give $\Delta^{w}(6)$.
Inset: expansion of the region around the zero intercept.}
\label{Fig:DOSexponentFE-ED}
\end{figure}

In order to stabilize the outer self-consistency cycle, 
we need an additional criterion.
To this end, we confirm numerically that
the density of states near the gap increases algebraically;
see~(\ref{dosexponenta}).
The weight $w_1=V_1^2$ of the peak in
$D_{\rm UHB}(\omega)$ at $\epsilon_1$
represents the integral of the density of states from
$\Delta(U)/2$ up to $\epsilon_1= \Delta(U)/2+
{\cal O}(1/n_s)$. Using~(\ref{dosexponenta})
we should find
\begin{equation}
\left[w_1\right]^{1/(\alpha+1)} = \epsilon_1-
\frac{\Delta^{w}(U)}{2}
\; .
\label{weightofns}
\label{weightcriterion}
\end{equation}
The linear behavior with $\alpha_{\rm FE-ED}=1/2$
is confirmed in Fig.~\ref{Fig:DOSexponentFE-ED}
for $U=6$. The plot also 
provides $\Delta^{w}(U)$ for a given $\Delta_{\rm in}(U)$
as the intersection of the extrapolated curves with the frequency
axis (`weight criterion').

The two values for the gap from the `energy criterion'~(\ref{energycriterion})
and from the `weight criterion'~(\ref{weightcriterion})
do not agree perfectly.
Therefore, we use the self-consistency condition
\begin{equation}
\Delta_{\rm in} (U) \stackrel{!}{=} \Delta_{\rm out}(U) \equiv 
\frac{\Delta^{E}(U)+\Delta^{w}(U)}{2} \; ,
\label{consistencycriterion}
\end{equation}
which determines $\Delta^{\rm FE-ED}(U)$ after convergence
of the outer self-consistency cycle.
The difference,
\begin{equation}
\delta\Delta^{\rm FE-ED}(U)=
\frac{|\Delta^{E}(U)-\Delta^{w}(U)|}{2}
\end{equation}
is our estimate for the accuracy of $\Delta^{\rm FE-ED}(U)$.
We call this implementation of the DMFT 
the `Fixed-Energy Exact Diagonalization (FE-ED)' scheme.

\begin{figure}[ht]
\resizebox{\columnwidth}{!}{%
\includegraphics{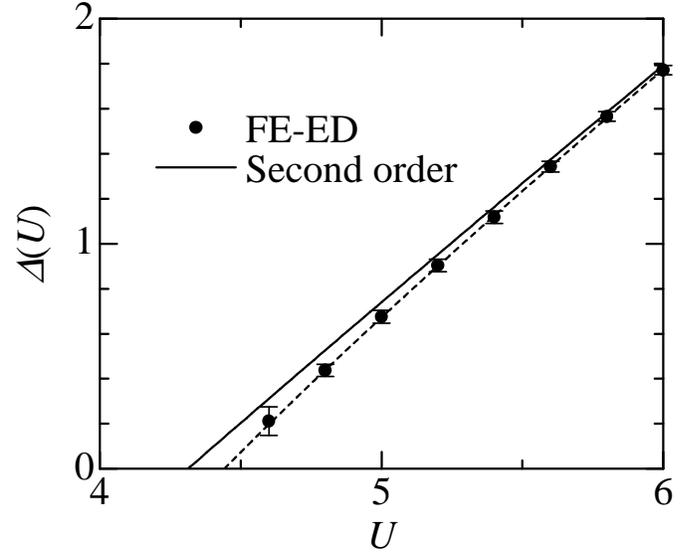}}
\caption{Mott--Hubbard gap as a function of $U$ in FE-ED in
comparison with the $1/U$ expansion to second
order~(\protect\ref{gapfinal3rd}).
The error bars give the two values which 
enter~(\protect\ref{consistencycriterion}) as obtained from the 
energy and weight criterion.
The dashed line is a cubic polynomial interpolation based on the data points
for $U\geq 4.8$.}
\label{Fig:FE-EDgap}
\end{figure}

The converged FE-ED gap as a function of the interaction strength
is shown in Fig.~\ref{Fig:FE-EDgap}. The agreement with the $1/U$~expansion
is excellent for $U\geq 5$ because 
the two curves agree within the error bars.
Unfortunately, the uncertainty 
in $\delta\Delta^{\rm FE-ED}(U)$
increases towards the transition. Using a cubic polynomial extrapolation
for data points $U\geq 4.8$, as shown in the figure, leads to 
$U_{\rm c}^{\rm FE-ED}= 4.44$.
As seen from the figure, the spline interpolation also touches
the data point for $U=4.6$, to within its error bars.
If this data point were included, the critical value 
would be $U_{\rm c}^{\rm FE-ED}= 4.42$.
Thus, we give
\begin{equation}
U_{\rm c}^{\rm FE-ED}= 4.43 \pm 0.05
\label{bestestimate}
\end{equation}
as our best estimate for the location of the closing of the gap.

\begin{figure}[ht]
\resizebox{\columnwidth}{!}{%
\includegraphics{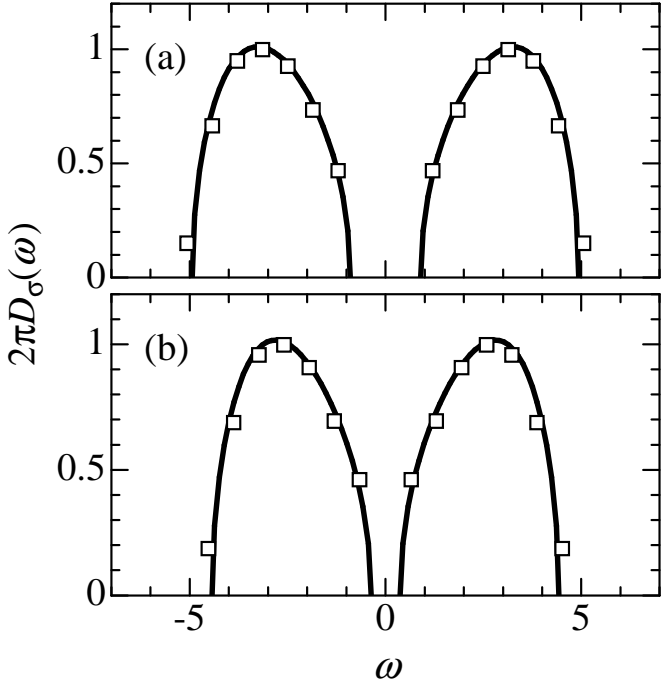}
}
\caption{Density of states, $D_{\sigma}(\omega)=D(\omega)/2$,
as obtained from the $1/U$ expansion,
eq.~(\protect\ref{DOSfinalres}), and the FE-ED for $n_s=15$;
(a) $U=6$, (b) $U=5$.}
\label{Fig:FE-EDDOS}
\end{figure}

In Fig.~\ref{Fig:FE-EDDOS} we compare the density of states of
our FE-ED for $n_s=15$ with those of the $1/U$ expansion
to second order. Both for $U=5$ and $U=6$ the agreement is 
very good. Our FE-ED thus allows us to determine 
the density of states quantitatively
with increasing resolution as $n_s$ increases.

Aside from the fact that the $1/U$ expansion
provides an educated guess for the position of
the Hubbard bands, the FE-ED is an independent
numerical method based on the DMFT. 
The good agreement of the results mutually supports the 
applicability of either method.
Therefore, we are confident that neither of the
limitations poses a serious problem for our investigation
of the Mott--Hubbard insulator, i.e., the problem of convergence
of the series expansions in $1/U$ and in $1/n_s$, respectively,
are not too serious in our approaches.

\section{Random Dispersion Approximation}
\label{sec:RDA}

In this section,
we present numerical results from the 
Random Dispersion Approximation (RDA). This method  
is not based on the DMFT and 
thereby provides yet another independent check
for the validity of our results as obtained in 
Sects.~\ref{sec:Strong} and~\ref{sec:FEED}.

In the Random Dispersion Approximation, the dispersion relation
$\epsilon(\sitek)$
in the kinetic energy is replaced by a random quantity 
$\epsilon^{\rm RDA}(\sitek)$
where the bare density of states acts as probability distribution,
\begin{equation}
\rho(\epsilon) = \frac{1}{L} 
\sum_{\sitek} \delta(\epsilon-\epsilon^{\rm RDA}(\sitek))
\; .
\label{dosdefRDA}
\end{equation}
This is the characteristic property of the dispersion relation in
infinite dimensions~\cite{Gebhardbook} so that the RDA 
with the semi-elliptic density of states~(\ref{rhozero})
becomes exact for the Bethe lattice with infinite coordination number
in the thermodynamic limit $L\to\infty$.

All correlation functions factorize, analogously to~(\ref{RDADq}).
For the $m$th-order
correlation function,
\begin{equation}
\rho_{1\ldots m} (\epsilon_1,\ldots,\epsilon_m) =
\frac{1}{L} \sum_{\sitek} \prod_{l=1}^m 
\delta\left(\epsilon_l-\epsilon(\sitek+\siteql)\right) \; ,
\end{equation}
we give a recursive algorithm. Let
$\rho_l(\epsilon_l)\equiv \rho(\epsilon_l)$ be the bare density of states,
for $m=2$, see~(\ref{RDADq}), and for $m\geq 3$ we define 
\begin{eqnarray}
\rho_{1\ldots (n-1);n,\ldots,m}(\epsilon_1,\ldots,\epsilon_m)
&=& \prod_{i,j=n}^m (1-\delta_{\siteqi,\siteqj}) \nonumber \\
&& 
\prod_{i=n}^m\prod_{j=1}^{n-1}  (1-\delta_{\siteqi,\siteqj})
\\
&& \rho_{1\ldots m} (\epsilon_1,\ldots,\epsilon_m)  \; .
\nonumber 
\end{eqnarray}
In this quantity, all $\siteql$ for $l\geq n$ are pairwise different,
and they are also different from those $\siteqi$ with $i\leq n-1$.
The recursion formula
\begin{eqnarray}
\lefteqn{\rho_{1\ldots n;(n+1),\ldots,m}(\epsilon_1,\ldots,\epsilon_m)=}
\nonumber \\
\lefteqn{\rho_{1\ldots (n-1);n,\ldots,m}(\epsilon_1,\ldots,\epsilon_m)}
\nonumber \\
&& + \left[ 1 - \prod_{l=1}^{n-1} \left(1- \delta_{\siteqn,\siteql}
\delta(\epsilon_n-\epsilon_l)\right) \right]
\\
&&\times \rho_{1\ldots (n-1);(n+1),\ldots,m}(\epsilon_1,\ldots,\epsilon_{n-1},
\epsilon_{n+1},\ldots,\epsilon_m)
\nonumber
\end{eqnarray}
starts with $\rho_{1\ldots m}(\epsilon_1,\ldots\epsilon_m)
\equiv \rho_{1\ldots m;}(\epsilon_1,\ldots\epsilon_m)$
and terminates at
\begin{equation}
\rho_{1;2,\ldots ,m}(\epsilon_1,\ldots\epsilon_m)
= \prod_{l=1}^m \rho(\epsilon_l) \prod_{i,j=1}^m (1-\delta_{\siteqi,\siteqj})
\; .
\end{equation}
For example,
\begin{eqnarray}
\rho_{123} (\epsilon_1,\epsilon_2,\epsilon_3)
&=&
\Bigl[ 1 - \prod_{l=1}^2 \left(1- \delta_{\siteq_3,\siteql}
\delta(\epsilon_3-\epsilon_l)\right) \Bigr]
\rho_{12}(\epsilon_1,\epsilon_2)\nonumber \\
&& + \rho_{12;3}(\epsilon_1,\epsilon_2,\epsilon_3)
\end{eqnarray}
with
\begin{eqnarray}
\rho_{12;3}(\epsilon_1,\epsilon_2,\epsilon_3) &=&
\delta_{\siteq_1,\siteq_2}\delta(\epsilon_1-\epsilon_2) 
\rho(\epsilon_1,\epsilon_3) \nonumber\\
&& + \rho(\epsilon_1)\rho(\epsilon_2)\rho_(\epsilon_3) \\
&& \times 
(1-\delta_{\siteq_1,\siteq_2})
(1-\delta_{\siteq_1,\siteq_3})
(1-\delta_{\siteq_2,\siteq_3}) \; .
\nonumber
\end{eqnarray}

In order to put the RDA into practice, we choose a one-dimensional 
lattice of $L$ sites in momentum space 
\begin{equation}
k_{\ell} = \frac{2\pi}{L} \left( -\frac{L+1}{2} +\ell \right) \;  
(\ell = 1,\ldots,L) 
\; ,
\end{equation}
and determine the dispersion relation
$\epsilon (k)$ as the solution of the implicit equation
\begin{equation}
k/2 = \left(2\epsilon(k)/W\right)
[1-\left(2\epsilon(k)/W\right)^2]^{1/2}
+\arcsin\left(2\epsilon(k)/W\right)
\; .
\end{equation}
This choice guarantees $\rho(\epsilon)=\rho_0(\epsilon)$
in the thermodynamic limit.

Next, we choose a permutation ${\cal Q}_{\sigma}$ for each
spin direction~$\sigma$ which permutes
the sequence $\{1,\ldots,L\}$ 
into $\{{\cal Q}_{\sigma}[1],\ldots,{\cal Q}_{\sigma}[L]\}$;
in this way there are $(L!)^2$ independent realizations
for a given finite~$L$.
A specific realization of the RDA dispersion is then defined by
${\cal Q}=[{\cal Q}_{\uparrow},{\cal Q}_{\downarrow}]$.
Then, the numerical task is the Lanczos diagonalization of
the Hamiltonian
\begin{equation}
\hat{H}^{{\cal Q}}
= \sum_{\sigma}\sum_{\ell=1}^{L} \epsilon(k_{{\cal Q}_{\sigma}[\ell]})
\hat{c}_{k_{\ell},\sigma}^+\hat{c}_{k_{\ell},\sigma}
 + U \hat{D}
\; .
\end{equation}
In this way we obtain 
the ground-state energy 
$E_0^{\cal Q}(N;U)$, the single-particle gap
\begin{equation}
\Delta^{\cal Q}(U)=2 [E_0^{\cal Q}(N=L+1;U)
-E_0^{\cal Q}(N=L;U)] \; ,
\label{rdagapformula}
\end{equation}
and the momentum distribution
\begin{equation}
n^{\cal Q}(\epsilon;U)=\frac{1}{2} \sum_{\sigma}
\left. \vphantom{\sum}
\langle \hat{n}_{k_{\ell},\sigma} \rangle
\right|_{\epsilon(k_{\ell})=\epsilon} \; ,
\end{equation}
where $\langle \ldots \rangle$ denotes the ground-state 
expectation value for the realization~${\cal Q}$.

\begin{figure}[ht]
\resizebox{\columnwidth}{!}{%
\includegraphics{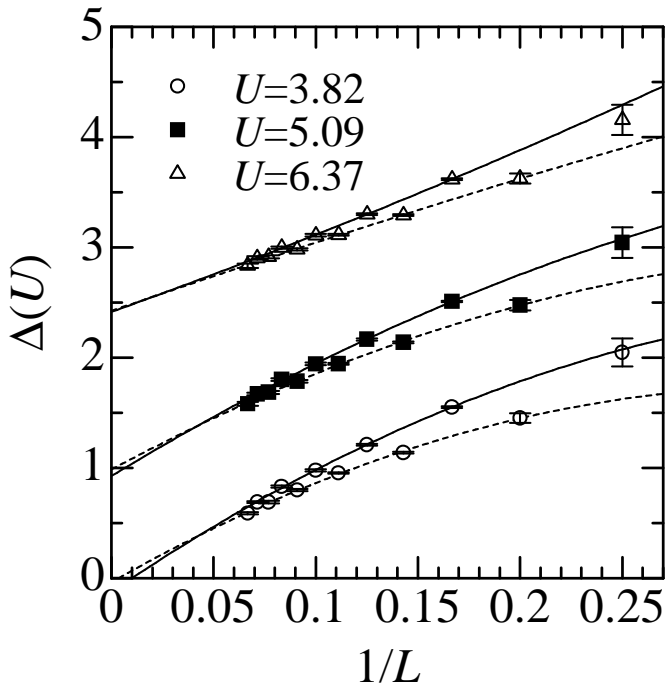}}
\caption{Gap as a function of system size ($L\leq 15$) in the Random
Dispersion Approximation for various values of the
interaction strength. The extrapolations are done for even and odd
system sizes separately.\label{fig:rdagapL}}
\end{figure}

As a next step, we obtain all physical quantities
for fixed system size~$L$ by averaging over
the number $N_{\cal Q}$ of realizations~${\cal Q}$.
Typically, we choose at least $N_{\cal Q}=100$ for
$6 \leq L\leq 16$.
For the physical quantities
we obtain gaussian-shaped distributions for which we
can determine the average values, e.g.,
\begin{eqnarray}
\Delta(U)&=&\frac{1}{N_{\cal Q}}\sum_{{\cal Q}} 
\Delta^{\cal Q}(U) \; , \\
n(\epsilon;U)&=&\frac{1}{N_{\cal Q}}\sum_{{\cal Q}} 
n^{\cal Q}(\epsilon;U)
\end{eqnarray}
with accuracy ${\cal O}(1/N_{\cal Q})$.

\begin{figure}[ht]
\resizebox{\columnwidth}{!}{%
\includegraphics{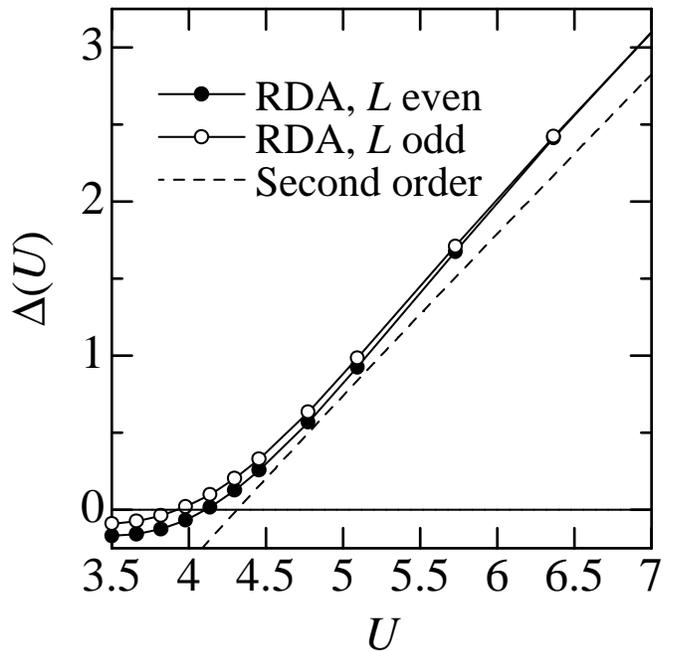}
}
\caption{Gap in RDA as a function of the interaction strength.
Filled symbols: extrapolation from even~$L$,
open symbols: extrapolation from odd~$L$. Also shown is
the result from second-order perturbation theory in 
$1/U$~(\ref{gapfinal3rd}).\label{fig:rdagap}}
\end{figure}

In order to improve slightly the quality of our distributions,
we impose a filter on our randomly chosen permutations. 
For a truly random dispersion, $L |t^{\rm RDA}(\ell)|^2 
=\overline{\epsilon^2}$ is independent
of~$\ell$; see~(\ref{defjq}). Therefore, we discard those
realizations for which 
\begin{equation}
\sum_{\ell=1}^{L-1} \left[|t^{{\cal Q}_{\sigma}}(\ell)|^2
- \overline{\epsilon^2} \right]^2 > d_L
\end{equation}
with, $d_L \approx 0.2$ for $L\leq 16$.
In this way we admit about every second
of the randomly chosen configurations.
Note that there are of the order of $(L!)^2$ 
different realizations so that our filter does not 
introduce any unwanted bias.

In Fig.~\ref{fig:rdagapL} we show the gap as 
calculated from~(\ref{rdagapformula}). Typically,
the statistical errors are 
of the symbol size or smaller. 
Due to the sizable odd-even effect,
we extrapolate the data for odd and even~$L$ separately.
This gives another check on the influence of finite-size effects.
It is seen that the behavior as a function of $1/L$ is quite
regular, i.e., the finite-size gap behaves as in
generic many-body problems. 

In Fig.~\ref{fig:rdagap} we show the result of an extrapolation
of the finite-size data for even and odd system sizes.
We consider the agreement with the $1/U$ expansion as quite acceptable
for $U>U_{\rm c}$ which we can estimate as
\begin{equation}
U_{\rm c}^{\rm RDA}= 4.0 \pm 0.4 \; .
\end{equation}
Finite-size effects result in a smearing of the gap at the transition,
and a gap exponent cannot be determined.
However, the fact that the extrapolated gap turns slightly negative can be
taken as an indication that the gap opens
at $U_{\rm c}$ with an exponent not too far from unity.

\begin{figure}[ht]
\resizebox{\columnwidth}{!}{%
\includegraphics{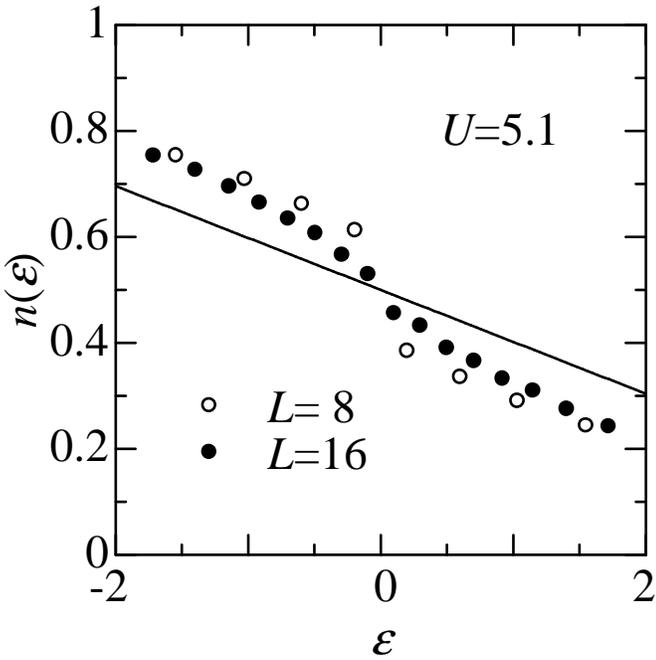}
}
\caption{Momentum distribution in RDA 
for $U=16/\pi\approx 5.1$ for the two largest system sizes.
The full line is the result from first-order perturbation theory 
in $1/U$~(\ref{nofkcoeff}).}
\label{Fig:rdamomentum}
\end{figure}

Lastly, in Fig.~\ref{Fig:rdamomentum} we show the momentum
dispersion within the RDA for $U=16/\pi\approx 5.1$, together with the
result of the $1/U$ expansion.
Obviously, there are large finite-size effects 
which make a finite-size extrapolation difficult. 
The reason for this behavior is related to the fact that
all finite-size systems appear to have a finite discontinuity
at the Fermi energy as well as a finite single-particle gap.
For insulators, the apparent jump in $n(\epsilon)$ only
vanishes in the thermodynamic limit~\cite{RDA}. The agreement
between the RDA data and the momentum distribution 
as obtained from the $1/U$ expansion is good enough to
argue that the RDA will eventually converge to the $1/U$ result
for $L\to\infty$.

The RDA becomes exact for lattice electrons in high dimensions.
In contrast to the FE-ED it is not based on the DMFT 
self-consistency equations. Moreover, it does not require
the convergence of the $1/U$ expansion.
Despite the observed limitations in accuracy, the results from RDA
confirm of our findings in Sects.~\ref{sec:Strong}
and~\ref{sec:FEED}: the $1/U$ expansion and the FE-ED
provide an accurate description of the Mott--Hubbard insulator
whose gap opens at the critical interaction strength
$U_{\rm c}=4.43 \pm 0.05$. 

\section{Comparison with analytic approximations}
\label{sec:companalyt}

In this section we compare our results with
two analytic approximations to the DMFT which become exact
for the Bethe lattice in the strong-coupling limit, i.e., they reproduce
$D_{\rm LHB}(\omega)$ to order $(1/U)^{0}$. 
The Local Moment Approach (LMA)~\cite{LMA}
outlined in appendix~\ref{AppendixB}
fulfills this condition by construction, whereas this is 
a non-trivial result 
within the Iterated Perturbation Theory (IPT)~\cite{iptoriginal}; 
a proof~\cite{Mikethesis} is given
in appendix~\ref{AppendixA}.

The Hubbard-III approximation~\cite{HubbardIII} also fulfills this criterion
but it fails to reproduce qualitatively the density of states 
for smaller interaction strengths~\cite{Eva}.
Therefore, we do not further discuss the Hubbard-III approximation.

\subsection{Iterated Perturbation Theory}
\label{subsec:IPT}

For the $Z\to\infty$ Bethe lattice, the self-energy
and the local single-particle Green function are related by  
\begin{equation}
\label{gipt}
G_{\sigma}(\omega)=\frac{1}{\omega-G_{\sigma}(\omega)
-\Sigma_{\sigma}(\omega)}.
\end{equation}
In IPT, the self-energy is approximated by
\begin{equation}
\Sigma_{\sigma}(\omega)=
U^2\int_{-\infty}^{\infty}\frac{\diff\Omega}{2\pi \I} 
\Pi(\Omega){\cal G}_{\sigma}(\omega-\Omega)\; .
\label{sipt}
\end{equation}
Here ${\cal G}_{\sigma}(\omega)$ is the host Green function, 
\begin{equation}
\label{scrgipt}
{\cal G}_{\sigma}(\omega)=\frac{1}{\omega-G_{\sigma}(\omega)},
\end{equation}
and we have defined the polarization bubble 
\begin{equation}
\label{piipt}
\Pi(\omega)= - \int_{-\infty}^{\infty}  \frac{\diff\omega_1}{2\pi\I}
{\cal G}_{\sigma}(\omega_1)
{\cal G}_{\sigma}(\omega_1-\omega)
\end{equation}
in terms of the host Green function.

It is helpful to consider the low-frequency
behavior of the IPT equations. 
First, consider the host Green function~(\ref{scrgipt}),
whose low-frequency behavior in the insulator is given generally by
\begin{equation}
\label{scrg0}
{\cal G}_{\sigma}(\omega)=\frac{\mod{m_0}}{\omega+\I\eta\, {\rm sgn}(\omega)}
\qquad \omega\to 0 \: ,
\end{equation}
where the pole weight $\mod{m_0}$ is given 
in terms of 
\begin{equation}
|\gamma_1|= -\left.\frac{\diff G_{\sigma}(\omega)}{\diff \omega}
\right|_{\omega=0}
\end{equation}
as
\begin{equation}
\label{mug1}
\mod{m_0}=(1+\left\vert\gamma_1\right\vert)^{-1}\; .
\end{equation}
${\cal G}_{\sigma}(\omega)$ also contains a Hubbard-band contribution, 
which is strictly separated from the pole at $\omega=0$ 
throughout the insulating phase, whence we may write 
\begin{equation}
\label{gband}
{\cal G}_{\sigma}(\omega)={\cal G}_{\sigma}^{\rm band}(\omega)
+\frac{\mod{m_0}}{\omega+\I\eta\, {\rm sgn}(\omega)}\; ,
\end{equation}
which actually defines ${\cal G}_{\sigma}^{\rm band}(\omega)$.   
In IPT the low-frequency behavior of 
${\cal G}_{\sigma}(\omega)$ directly
determines the low-frequency behavior of $\Sigma_{\sigma}(\omega)$; 
inserting~(\ref{gband}) into~(\ref{sipt}) and~(\ref{piipt}) yields
\begin{equation}
\label{siptlow}
\Sigma_{\sigma}(\omega)=\frac{A}{\omega}\qquad \omega\to 0 \: ,
\end{equation}
where 
\begin{equation}
\label{Amu}
A=\frac{U^2}{4}\mod{m_0}^3\; .
\end{equation}
To satisfy self-consistency~(\ref{gipt}) requires
\begin{equation}
\label{Ag1}
A=\frac{1}{\left\vert\gamma_1\right\vert}\; ,
\end{equation}
so that (\ref{mug1}), (\ref{Amu}) and~(\ref{Ag1}) determine the low-fre\-quency
behavior of the problem. In particular, they yield
\begin{equation}
\label{muU}
\frac{2}{\mod{m_0}\sqrt{1-\mod{m_0}}}=U\; .
\end{equation}
The denominator
${\mod{m_0}\sqrt{1-\mod{m_0}}}$ reaches its maximum value when
\begin{equation}
\label{mumin}
\mod{m_0}=\mod{m_0^{\rm min}}=\frac{2}{3}\; ,
\end{equation}
thus locating the destruction of the Mott--Hubbard insulator
at
\begin{equation}
\label{Umin}
U_{{\rm c},1}^{\rm IPT}=3\sqrt{3}\approx 5.2 \; ,
\end{equation}
as earlier reported by Rozenberg et al.~\cite{KotliarIPT}.
The IPT prediction
lies considerably above our best estimate $U_{\rm c}= 4.43 \pm 0.05$.

\begin{figure}[htb]
\resizebox{\columnwidth}{!}{%
\includegraphics{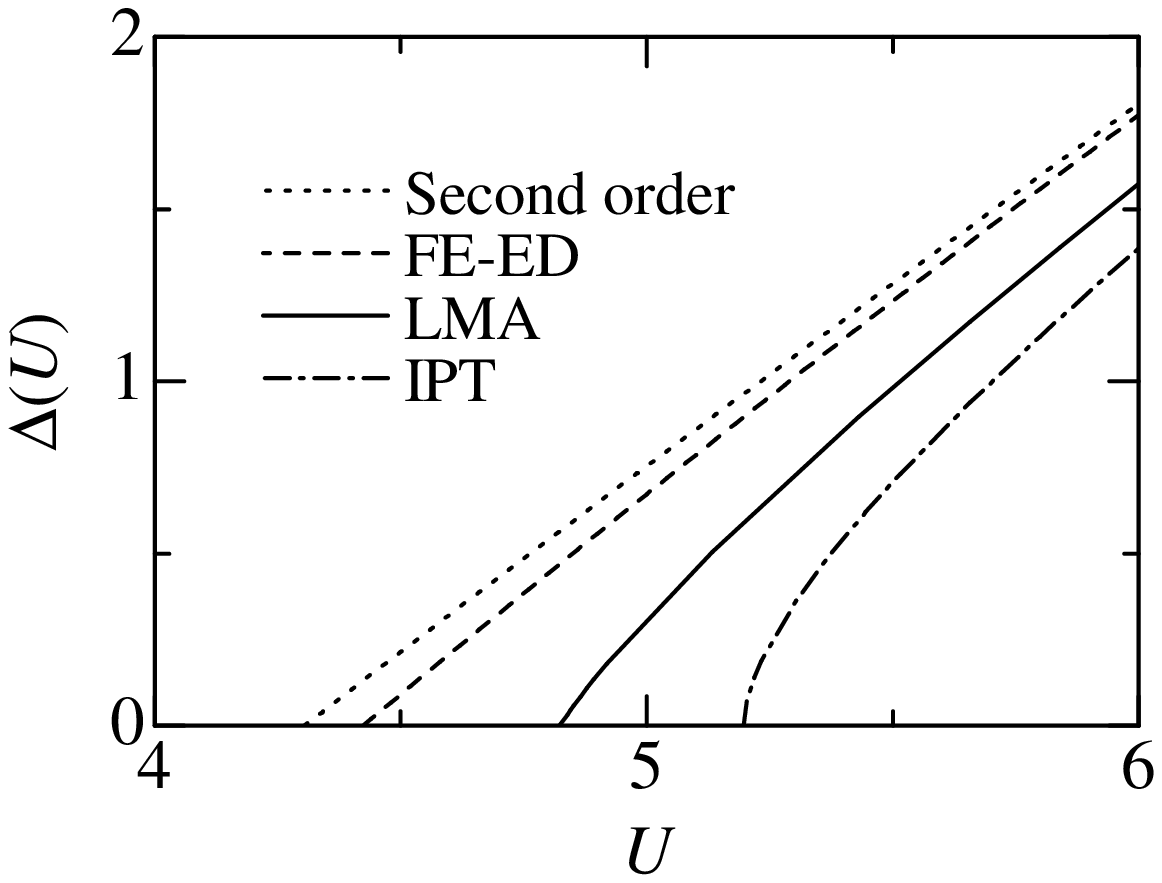}
}
\caption{Gap as a function of interaction strength
from second-order perturbation theory 
in~$1/U$~(\protect\ref{gapfinal3rd}), the
Fixed-Energy Exact Diagonalization (FE-ED), 
the Local Moment Approach (LMA), and Iterated Perturbation Theory (IPT).
\label{fig:LMAgap}\label{fig:iptgap}}
\vspace{20pt}
\resizebox{\columnwidth}{!}{%
\includegraphics{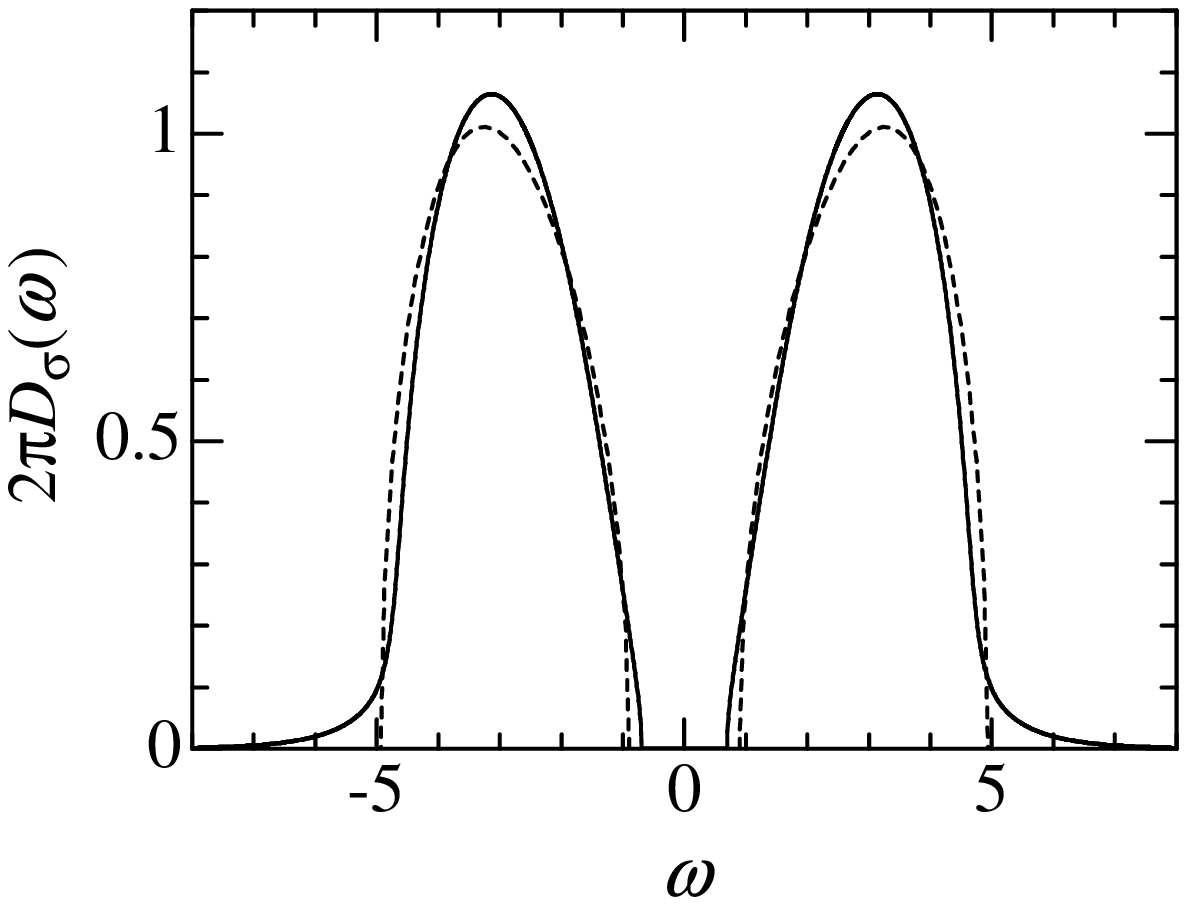}}
\caption{Density of states for $U=6$ in Iterated Perturbation Theory 
(solid line) and in second-order perturbation theory 
in~$1/U$~(\protect\ref{DOSfinalres}) (dashed line).
\label{Fig:IPTDOS}}
\end{figure}

This discrepancy can already be seen for larger interaction strength
where
\begin{equation}
\Delta^{\rm IPT}(U)= U -4 - \frac{5}{2U} - \ldots
\; .
\end{equation}
The first-order correction can be obtained analytically,
as shown in appendix~\ref{AppendixA}.
It is more than twice as large as 
our exact coefficient. 
Therefore, the agreement is good only for $U>6$. 
The gap in IPT is compared to our FE-ED and second-order results in
Fig.~\ref{fig:iptgap}.
As seen from the figure, and shown analytically in~\cite{Mikethesis},
the gap closes continuously but with an 
exponent of $\gamma_{\rm IPT}=1/2$.

Despite the deviations in the gap values,
the IPT correctly reproduces the overall shape
of the density of states for $U>6$, apart from
the artificial high-energy tails of the IPT Hubbard bands.
The threshold exponent for the density of states is correct,
$\alpha_{\rm IPT}=1/2$ in the Mott--Hubbard insulator.
Nevertheless, the deviations from the 
second-order result~(\ref{DOSfinalres}) are quite noticeable at $U=6$
which is shown in Fig.~\ref{Fig:IPTDOS}. 
We conclude that the IPT provides a 
quantitatively correct description
of the Mott--Hubbard insulator down to $U\approx 6$.
However, IPT seriously underestimates the stability of the
Mott--Hubbard insulator.

\subsection{Local Moment Approach}
\label{subsec:LMA}

Details of the LMA are given in Ref.~\cite{LMA}. 
As shown in appendix~\ref{AppendixB}, the LMA gap 
to first order in $1/U$ is given by
\begin{equation}
\Delta^{\rm LMA}(U)= U - 4 - \frac{3}{2U} 
\; .
\end{equation}
The first-order correction is close to the exact one, and,
correspondingly, the agreement with our strong-coupling 
result~(\ref{gapfinal3rd})
is very good down to $U\approx 5.5$.
For smaller interaction strengths,
the deviations become noticeable and 
the LMA predicts the collapse
of the insulator to occur at $U_{\rm c}^{\rm LMA}= 4.82$~\cite{LMA} 
which is larger than
our best estimate $U_{\rm c}= 4.43 \pm 0.05$.
The LMA yields
a critical gap exponent $\gamma_{\rm LMA}=1$~\cite{Mikethesis},
in contrast to IPT where $\gamma_{\rm IPT}=1/2$.
The gap in LMA and our second-order and FE-ED results are shown in
Fig.~\ref{fig:LMAgap}. 

\begin{figure}[ht]
\resizebox{\columnwidth}{!}{%
\includegraphics{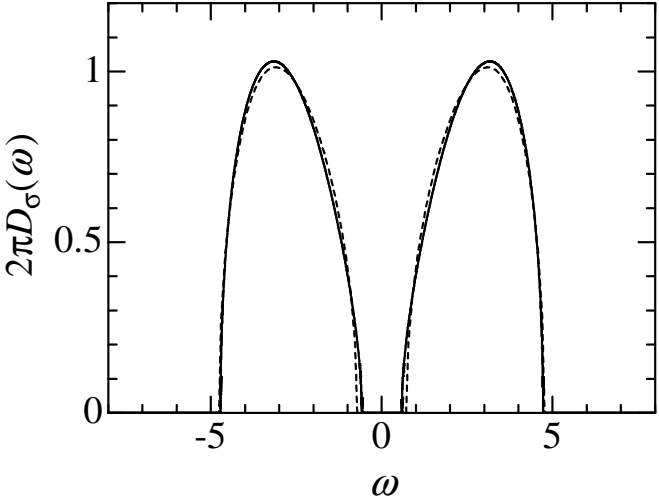}
}
\caption{Density of states for $U=4\sqrt{2}\approx 5.66$ 
from the Local Moment Approach (full line)
and second-order perturbation theory 
in~$1/U$~(\protect\ref{DOSfinalres}) (dashed line).
On the scale of the figure, the gap from FE-ED and from second-order
perturbation theory are almost the same.}
\label{Fig:LMADOS}
\end{figure}

It is seen that the LMA gap is more accurate than the gap from IPT.
In particular, this holds true for 
the density of states as the LMA very well reproduces our
second-order result~(\ref{DOSfinalres}). 
The comparison for $U=4\sqrt{2}\approx 5.66$ is shown in
Fig.~\ref{Fig:LMADOS}. The deviations are largest
around the gap. The overall agreement, however, is excellent.
In particular, the threshold exponent is $\alpha_{\rm LMA}=1/2$.

The LMA thus provides a qualitatively correct description
of the Mott--Hubbard insulator. It is
a quantitatively reliable approximation down to $U\approx 5.5$.

\section{Other numerical approximations to the DMFT}
\label{sec:garbagenumerics}

In this section we discuss two other exact-diagonalization approaches.
In both of them a discretized version of the 
single-impurity Anderson model (SIAM), 
Sect.~\ref{subsec:DMFT}, is investigated
in which the site energies $\epsilon_{\ell}$
are determined self-consistently.
This is the decisive difference to our `fixed-energy' algorithm
as presented in Sect.~\ref{sec:FEED}.

In this work, we do not discuss the Numerical Renormalization Group (NRG) 
approach to the Mott--Hubbard insulator.
The NRG is custom-tailored for the description
of the metallic phase in the sense that it tries to resolve structures
around $\omega=0$ whereas features at high energies are 
broadened on a logarithmic scale~\cite{BullaPRL}. Therefore,
the Mott--Hubbard insulator does not display a clear gap. 
Consequently,
as can be seen from Fig.~2 in Ref.~\cite{BullaPRL},
the density of states is finite in the gap,
and the NRG Hubbard bands display sizable high-energy tails.
Therefore, the gap and the width of the Hubbard bands
as a function of the interaction strength 
cannot be easily deduced from this approach.

\subsection{Exact diagonalization in `two-chain geometry'}
\label{subsec:ED1}

In contrast to Sect.~\ref{subsec:DMFT},
an alternative formulation of 
the Hamilton operator for the SIAM describes
two chains of length $(n_s-1)/2$ 
which represent the upper and lower
Hubbard bands. They are coupled to the impurity site at the respective
chain origins~\cite{RMP,Rozenbergetal}
\begin{eqnarray}
\hat{H}_{\rm SIAM-tc} &=&\sum_{\ell=1;\sigma}^{(n_s-1)/2}
\epsilon_{\ell} \left(\hat{u}_{\sigma;\ell}^+\hat{u}_{\sigma;\ell}
- \hat{l}_{\sigma;\ell}^+\hat{l}_{\sigma;\ell}\right)
\nonumber \\
&&+ U \left( \hat{d}_{\uparrow}^+\hat{d}_{\uparrow} -\frac{1}{2}\right)
\left( \hat{d}_{\downarrow}^+\hat{d}_{\downarrow} -\frac{1}{2}\right) 
\nonumber \\
&& + \sum_{\sigma} \sqrt{\frac{1}{2}} 
\left( \hat{u}_{\sigma;1}^+\hat{d}_{\sigma}
+ \hat{l}_{\sigma;1}^+\hat{d}_{\sigma} + {\rm h.c.}\right)
\label{SIAMED1ns}
\\
&& + \sum_{\ell=1}^{(n_s-3)/2} \!\!\!V_{\ell}
\left( \hat{u}_{\sigma;\ell+1}^+\hat{u}_{\sigma;\ell} + 
\hat{l}_{\sigma;\ell+1}^+\hat{l}_{\sigma;\ell} + {\rm h.c.}
\right) \; .
\nonumber
\end{eqnarray}
Here particle-hole symmetry results in
$\epsilon_{\ell}>0$ for $\ell=1,\ldots,(n_s-1)/2$, and $V_{\ell}>0$ 
for $\ell=1,\ldots,(n_s-3)/2$.

The main advantage of the two-chain geometry lies in the fact
that on the Bethe lattice 
the self-consistency condition~(\ref{selfcons}) can be
implemented exactly even for finite $n_s$. 
In the continued-fraction expansion of the Green function for the 
Hubbard bands~(\ref{DefGLHB}), the local energies $\epsilon_{\ell}$ 
and the electron transfer amplitudes $V_{\ell}$ appear as
diagonal and off-diagonal coefficients,
\begin{equation}
G_{\rm LHB;\sigma}(\omega) =
\frac{1/2}{\omega+U/2-(V_1)^2/[\omega-\epsilon_1 - \ldots]}
 \; .
\label{contfrac}
\end{equation}
The (dynamical) Lanczos procedure directly provides
the continued-fraction coefficients so that the output of
the exact diagonalization procedure yields
the input for the next iteration of the self-consistency
cycle without a collection of weights as, e.g., 
in~(\ref{recollectionofweight}).

\begin{figure}[ht]
\resizebox{\columnwidth}{!}{%
\includegraphics{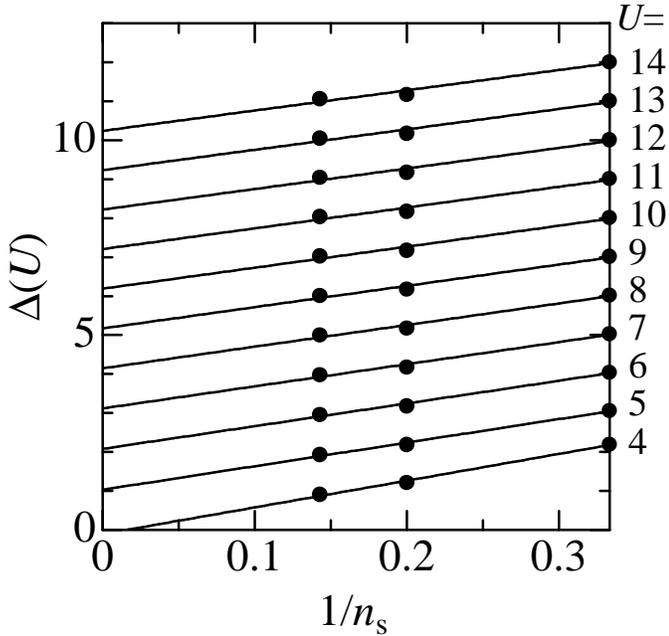}}
\caption{Gap in exact diagonalization of the 
single-impurity model in two-chain geometry 
for $n_s=3,5,7$ for various interactions~$U$.
The lines are a linear extrapolation to $1/n_s\to 0$.
}
\label{Fig:gapED1ns_small}
\end{figure}

\begin{figure}[ht]
\resizebox{\columnwidth}{!}{%
\includegraphics{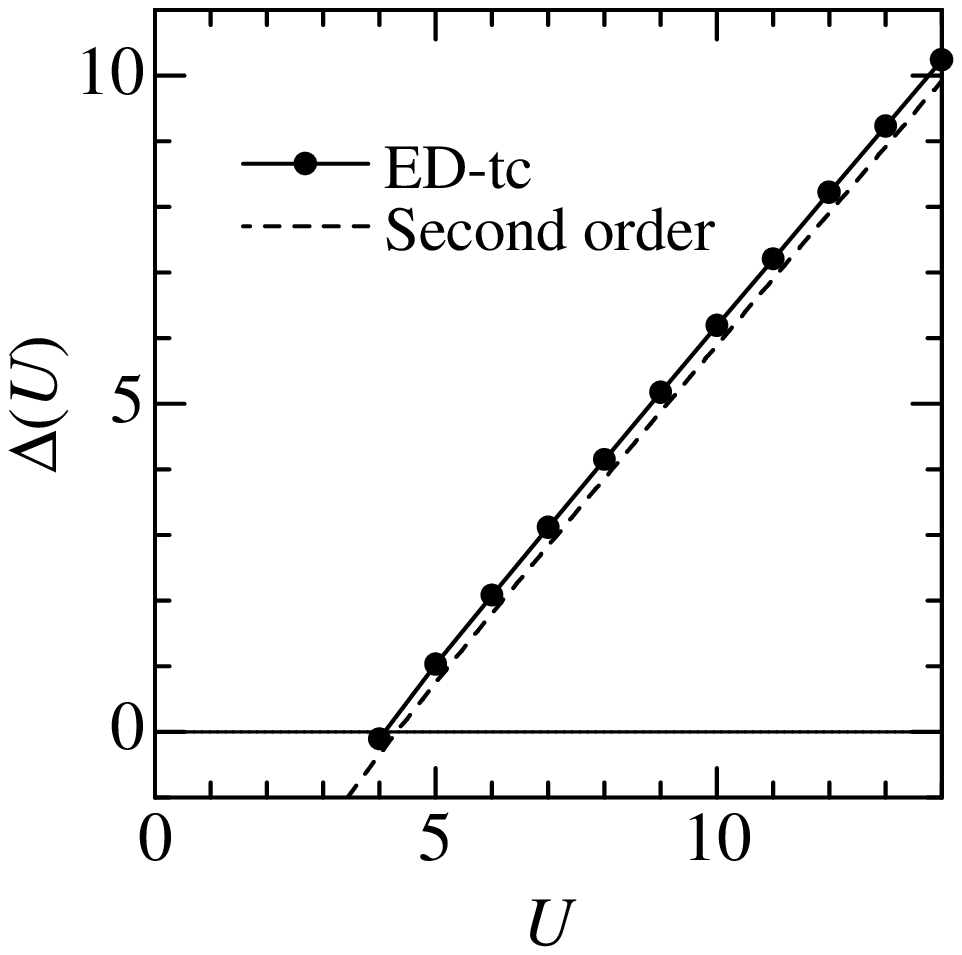}}
\caption{Extrapolated gap (see Fig.~\protect\ref{Fig:gapED1ns_small})
in the exact diagonalization of the 
single-impurity model in two-chain geometry 
as a function of the interaction strength,
as compared to second-order perturbation theory 
in~$1/U$~(\protect\ref{gapfinal3rd}).}
\label{Fig:gapED1_small}
\end{figure}

For this reason, the two-chain ED is rather appealing.
Moreover, using the results for $n_s=3,5,7$ and some 
unspecified extrapolation in $1/n_s$,
the gap in~\cite{RMP} was found to be almost linear as a function
of the interaction strength above $\widetilde{U}_{\rm c} =4.30$,
in surprisingly good agreement with our best estimate~(\ref{bestestimate}).
The corresponding results for $n_s=3,5,7$ are 
shown in Fig.~\ref{Fig:gapED1ns_small}.
If we take only those three data points into account and
assume a linear scaling in $1/n_s$, 
we obtain the gap as shown in Fig.~\ref{Fig:gapED1_small} which is
very similar to the result reported in~\cite{RMP},
and is in reasonable agreement with the results from the $1/U$ expansion.
Thus, one would be tempted to conclude that the method is quite
reliable. Unfortunately, such a conclusion is not warranted.

\begin{figure}[ht]
\resizebox{\columnwidth}{!}{%
\includegraphics{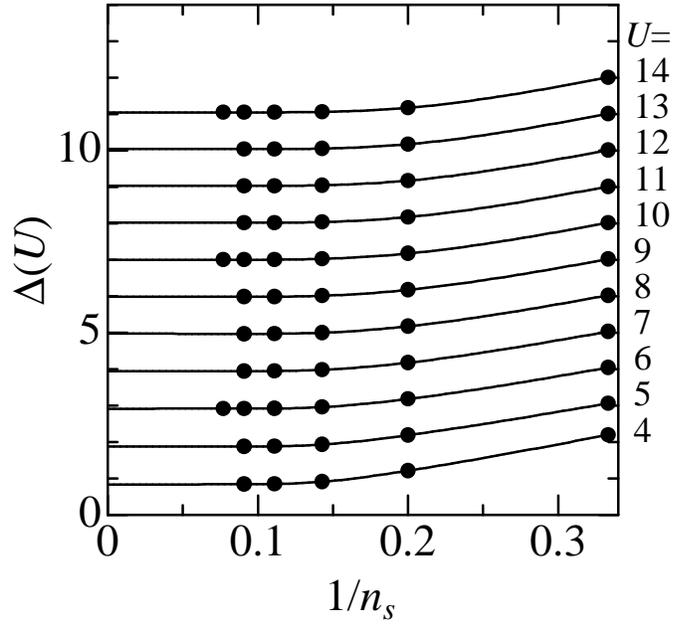}}
\caption{Gap in exact diagonalization of the 
single-impurity model in two-chain geometry for $n_s\leq 13$ 
for various interactions~$U$.
The lines are a cubic spline extrapolation to $1/n_s\to 0$.}
\label{Fig:gapED1ns_large}
\end{figure}

\begin{figure}[ht]
\resizebox{\columnwidth}{!}{%
\includegraphics{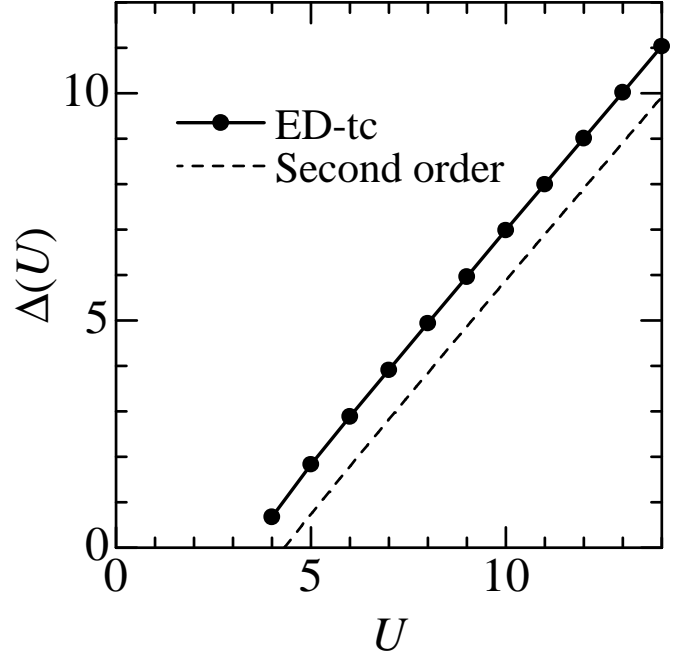}}
\caption{Extrapolated gap (see Fig.~\protect\ref{Fig:gapED1ns_large})
in the exact diagonalization of the 
single-impurity model in two-chain geometry 
as a function of the interaction strength,
as compared to second-order perturbation theory 
in~$1/U$~(\protect\ref{gapfinal3rd}).\label{Fig:gapED1_large}}
\end{figure}

First, bigger system sizes need to be analyzed because $n_s=3,5,7$
correspond to chain lengths of $L_{\rm c}=1,2,3$. When we follow the
original algorithm as described in~\cite{Rozenbergetal}, we find that
the insulating phase becomes unstable at $U=8$ for $n_s=13$
($U=6.4, 5.0, 3.5$ for $n_s=11,9,7$).
Obviously, this is a numerical artifact which is not related 
to the metal-insulator transition. We have found that 
this problem can be cured by incorporating the particle-hole symmetry
into $\hat{H}_{\rm SIAM-tc}$~(\ref{SIAMED1ns}) right from the start;
this was done only `on average' in Ref.~\cite{Rozenbergetal}.
In the following, we show results obtained using this particle-hole
symmetry.

When we include bigger system sizes,
we obtain a behavior quite different from the one
seen in Fig.~\ref{Fig:gapED1_small}.
The result for the gap as a function of $1/n_s$ for
$n\leq 13$ is shown in Fig.~\ref{Fig:gapED1ns_large}.
The behavior of the gap as a function of $1/n_s$
can no longer be approximated by a linear fit. 
The gap is not regular as a function of $1/n_s$ but seems to saturate. 
When we interpolate using a cubic spline, 
we obtain the gap as a function of the interaction strength
as shown in Fig.~\ref{Fig:gapED1_large}.
Obviously, the results are rather poor as
the large-$U$ limit cannot be reproduced properly
and there is nowhere agreement with the results 
from the $1/U$ expansion.
It is thus seen that the agreement 
between the data reported in~\cite{RMP,andere}
and ours is fortuitous.


\begin{figure}[ht]
\resizebox{\columnwidth}{!}{%
\includegraphics{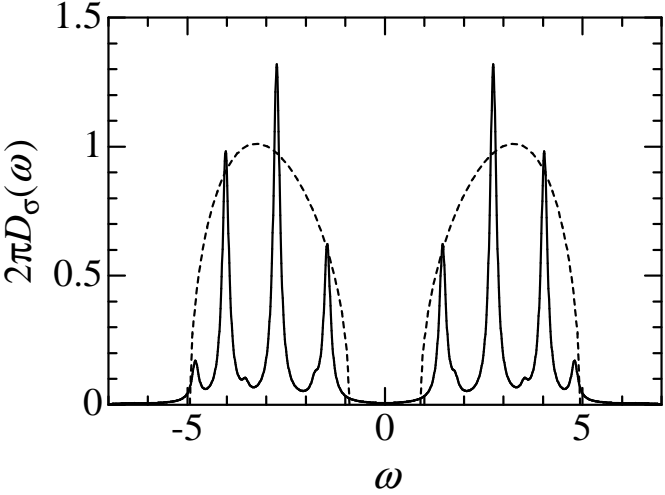}}
\caption{Density of states as obtained from the two-chain exact 
diagonalization 
for $U=6$ and $n_s=13$ (full line), 
as compared to second-order perturbation theory 
in~$1/U$~(\protect\ref{DOSfinalres}) (dashed line).\label{Fig:ED1DOS}}
\end{figure}

The poor quality of the two-chain ED is also seen
in the density of states, 
as shown in Fig.~\ref{Fig:ED1DOS} for $n_s=13$ and $U=6$.
For all $n_s\geq 7$ we obtain a four-peak structure
which does not allow the recovery of the true density of states.

The reason for the failure of the method lies in the setup of the
self-consistency scheme. In the continued-fraction expansion one
optimizes the moments of the density of states. 
However, the reconstruction of the density of states from
its moments is a numerically very delicate inverse problem.
In our case, the moments of the upper Hubbard band 
can be approximated very well by three main peaks 
and many small peaks at (much) higher energy. For this reason,
the structure of the density of states is essentially unchanged
as a function of $n_s$. The convergence as a function of system size
saturates, and a reliable extrapolation scheme is not evident. 
We thus conclude that the two-chain ED is
not a suitable tool for the investigation
of the Mott--Hubbard insulator.

\subsection{Caffarel--Krauth exact diagonalization in `star geometry'}
\label{subsec:OnoED2}

As an alternative to the two-chain discretization
of the previous section, Caffarel and Krauth
used the discretization of the single-impurity 
Anderson model in `star geometry'~(\ref{SIAMns}).
In contrast to our FE-ED, both the hybridization
parameters~$V_{\ell}$ and the site energies $\epsilon_{\ell}$
are determined self-consistently in
the Caffarel--Krauth scheme (CK-ED). 

In order to close the self-consistency cycle, a
$\chi^2$-fitting procedure~\cite{CK,Ono} is used.
Let $T$ be a fictitious small temperature, typically $T=1/200$,
and $\omega_n=(2n+1)\pi T$ are the corresponding Matsubara frequencies.
For a given parameter set $(\epsilon_{\ell},V_{\ell})$
we solve the single-impurity model for the single-particle
Green function $G_{\sigma}^{(n_s)}(\omega)$.
Then, the self-consistency equation~(\ref{selfcons}) allows us to 
deduce the parameter set for the next iteration
from the minimization of
\begin{eqnarray}
\chi^2(\epsilon_{\ell}^{\rm new},V_{\ell}^{\rm new}) &=&
\sum_{n=0}^{n_{\rm M}} \Bigl| 
H^{(n_s)}(\I \omega_n;\epsilon_{\ell}^{\rm new},V_{\ell}^{\rm new})
\nonumber \\
&& \hphantom{\sum_{n=0}^{n_{\rm M}} \Bigl|}
- G_{\sigma}^{(n_s)}(\I \omega_n;\epsilon_{\ell},V_{\ell})
\Bigr|^2 \; .
\label{chsqfit}
\end{eqnarray}
Here $n_{\rm M}$ is a (large) upper cut-off.

\begin{figure}[ht]
\resizebox{\columnwidth}{!}{%
\includegraphics{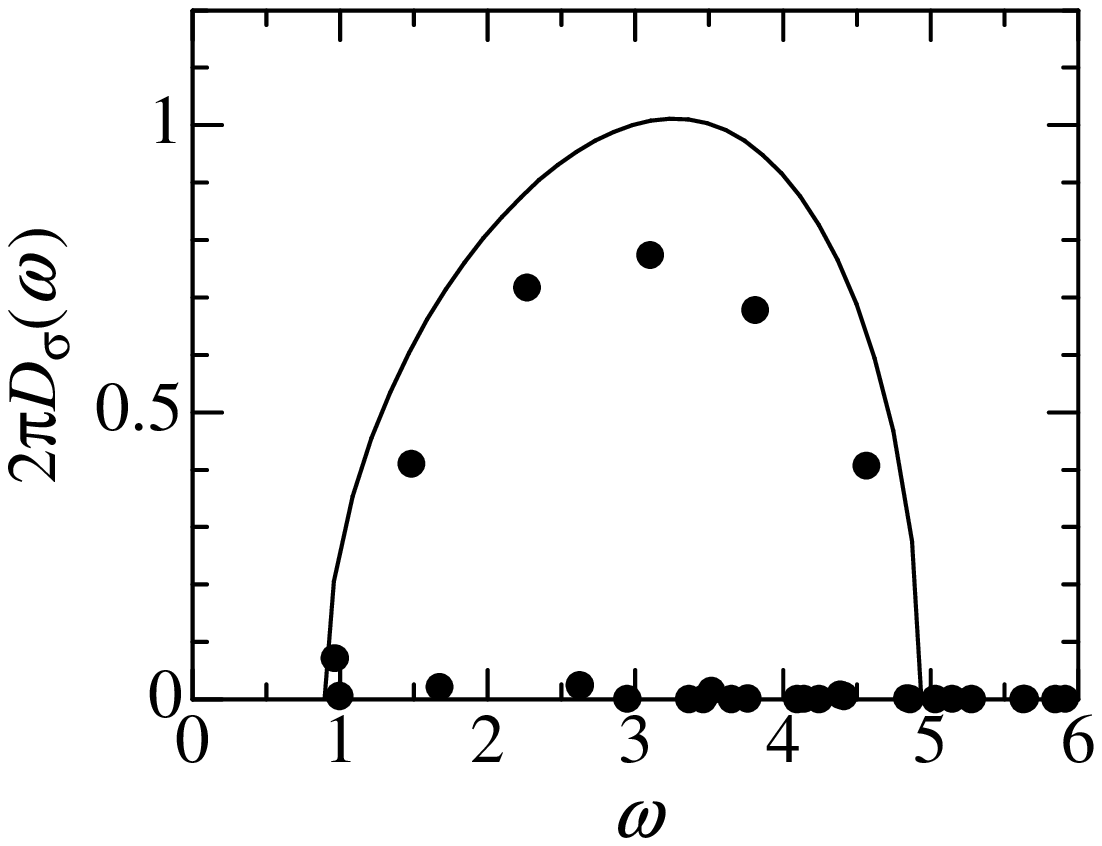}}

\vspace{3pt}

\resizebox{\columnwidth}{!}{%
\includegraphics{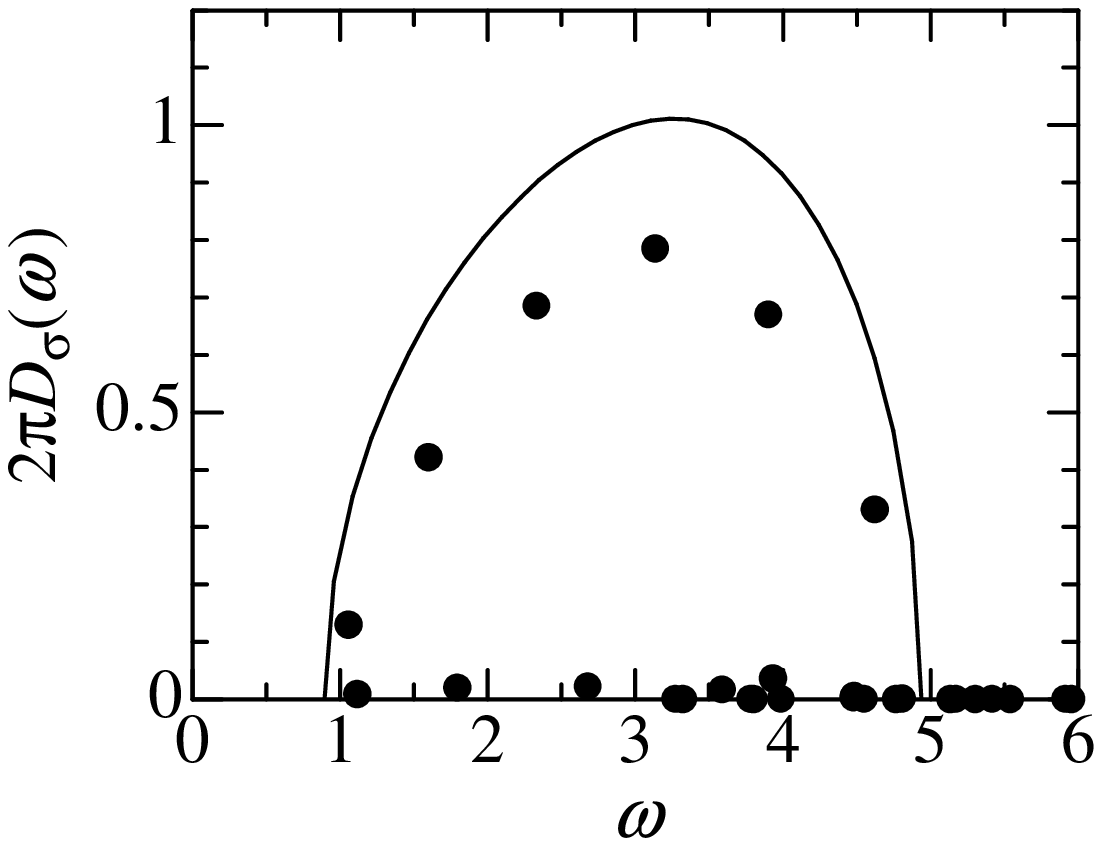}}
\caption{Density of states for $U=6$ as obtained from the $1/U$ expansion
(solid line), in comparison with two converged solutions of
the Caffarel--Krauth Exact Diagonalization scheme (CK-ED)
for $n_s=11$ sites (solid points).
\label{Fig:CK-EDdos}}
\end{figure}

In Fig.~\ref{Fig:CK-EDdos} we show two typical converged results for 
the density of states for $U=6$, starting from two
different initial values for the set $(V_{\ell},\epsilon_{\ell})$. 
Note that the solutions are not unique, i.e.,
there are many self-consistent solutions for the insulator.
At first glance, the two solutions appear to be very similar.
The main peaks have almost the same weights $V_{\ell}^2$ and the same 
positions $\epsilon_{\ell}$. This can be expected because the main
peaks dominate $\chi^2$ in~(\ref{chsqfit}). Moreover, the overall
distribution of weights and energies by and large 
follows the true density of states from the analytical calculation.

A closer look into the figures reveals a fundamental problem.
In Fig.~\ref{Fig:CK-EDdos}a the first non-vanishing peak appears
at $\Delta_a^{(11)}(6)/2=0.966$ whereas in
Fig.~\ref{Fig:CK-EDdos}b the value for half the gap
is estimated as $\Delta_b^{(11)}(6)/2=1.06$.
The same results follow if we use the difference of
the ground-state energies~(\ref{FE-EDgap}) for the definition
of the gap. Obviously, we cannot give a unique value
for $\Delta^{{\rm CK-ED};(n_s)}$, and an extrapolation
$n_s\to\infty$ is impossible.
The reason for this problem lies in the fact that 
the peak at the onset of the density of states is rather small 
so that its precise position and weight do not matter much
in the $\chi^2$-fit~(\ref{chsqfit}).
Thus, the onset energy may easily fluctuate by 10\% 
or more, as seen in Fig.~\ref{Fig:CK-EDdos}. Therefore,
the CK-ED approach cannot provide a reliable estimate for the gap.

The failure of the self-consistency can be traced back
to the fact that, for general filling, the Green function $G^{(n_s)}(\omega)$
contains only $n_s$ poles with substantial weight
whereas $H^{(n_s)}(\omega)$ is characterized by $(2n_s)$ parameters
$(\epsilon_{\ell},V_{\ell}^2)$ in the CK-ED. Too much flexibility
in the hybridization function leads to non-unique solutions
of the self-consistency equations. Our Fixed-Energy Exact Diagonalization
cures this problem because only $n_s$ parameters $V_{\ell}^2$
are to be determined.

\section{Conclusions}
\label{sec:conclusions}

In this work, we have studied the insulating phase of the
half-filled Hubbard model on a Bethe lattice with
infinite connectivity (band-width $W=4t$). 
As long as the strong-coupling perturbation theory
converges, the ground state of the Mott--Hubbard insulator has
a finite entropy density, $s=\ln(2)$, and the density of states
of the lower and upper Hubbard bands increases as a function
of frequency with the edge exponent $\alpha=1/2$, see~(\ref{dosexponentb}).
Furthermore, we find that the high-order corrections to the gap
as a function of $1/U$ are fairly small. 
Thus, the gap opens at the critical
interaction strength $U_{\rm c}^{\rm sc}=4.40 \pm 0.09$.
Our explicit results to second order 
in $1/U$~(\ref{DOSfinalres})
provide a benchmark test for analytical and numerical
approaches to the Mott--Hubbard insulator in infinite
dimensions. 

We have used our results from the $1/U$ expansion to set up
a new numerical scheme for the effective single-impurity Anderson model
in the Dynamical Mean-Field Theory. In our Fixed-Energy
Exact Diagonalization (FE-ED), we start an outer self-consistency cycle
with the position and width of
the Hubbard bands from perturbation theory which we
resolve equidistantly to accuracy $9t/(n_s-1)$ for $n_s \leq 15$. 
The extrapolation $n_s\to\infty$ thereby
becomes systematic.
In an inner self-consistency cycle we determine the 
hybridization strengths for given $n_s$ 
using a dynamical Lanczos procedure. From this we determine
the gap after extrapolation to the thermodynamic limit.
We merge the extrapolated gap from the `energy criterion' 
and the `weight criterion' in order to obtain a new estimate for the
onset of the Hubbard bands. In this way, we iterate to convergence
the outer self-consistency cycle.

Our FE-ED very well reproduces the gap, the density of states, and
the exponent $\alpha=1/2$ from
perturbation theory. In this way we confirm our estimate
for the transition to $U_{\rm c}^{\rm FE-ED}=4.43 \pm 0.05$.
As a last check, we favorably compare our results with those from
the Random Dispersion Approximation which is an independent
approach to the limit of infinite dimensions.
Thus, we are confident that the strong-coupling perturbation theory,
our Fixed-Energy Exact Diagonalization scheme for the
Dynamical Mean-Field Theory, and the Random Dispersion 
Approximation provide a consistent and accurate description
of the Mott--Hubbard insulator.

Other analytical and numerical techniques for the
solution of the Dynamical Mean-Field Theory meet
our benchmark test with limited success. The best
analytical approximation is the Local Moment Approach which
is quantitatively reliable for the density of states 
down to $U=5.5$; it reproduces the correct exponent $\alpha=1/2$
but overestimates the critical
interaction strength, $U_{\rm c}^{\rm LMA}=4.8$. 
Iterated Perturbation Theory
reproduces the correct threshold exponent $\alpha=1/2$
but it becomes quantitatively and qualitatively unreliable below $U=6$
and underestimates the stability of the Mott--Hubbard insulator,
$U_{\rm c}^{\rm IPT}=5.2$. 

Of the two exact-diagonalization (ED) schemes for the effective single-impurity
problem existing in the literature, the two-chain approach fails
because it leads to an ill-conditioned inverse
problem, namely, the recovery of the density of states from
its moments. The Caffarel--Krauth exact diagonalization
approach (CK-ED) in star geometry
gives a reasonable guess for the shape of the density of states,
but it fails
to provide a reliable estimate for the gap. 
For fixed system size, the fitting
procedure results in different solutions 
with noticeable differences for the onset of the density of states.
Therefore, it is not possible to set up a sensible extrapolation
scheme for $n_s\to\infty$. 

In this work we have presented a qualitatively and
quantitatively reliable analysis of the
Mott--Hubbard insulator. It will provide a solid basis
for further studies of the Mott--Hubbard metal-insulator
transition.

\subsubsection*{Acknowledgments}

We thank Eric Jeckelmann and David Logan for helpful discussions.
Support by the Deutsche Forschungsgemeinschaft
(GE 746/5-1+2) and the center {\sl Optodynamics\/}
of the Philipps-Universit\"at Marburg is gratefully acknowledged.
We thank the HRZ Darmstadt computer facilities where some
of the calculations were performed.

\appendix

\section{Iterated Perturbation Theory (IPT) in strong coupling}
\label{AppendixA}

We here write the IPT self-energy, cf.~(\ref{gipt})--(\ref{piipt}),
in a form amenable to numerical
calculations and additionally find its strong coupling form. 
We start by
rearranging~(\ref{gband}) as follows
\begin{eqnarray}
\nonumber
{\cal G}_{\sigma}(\omega)&=&{\cal G}_{\sigma}^{\rm band}(\omega)
+\frac{\mod{m_0}}{\omega+\I\eta\, \sgn(\omega)}\\
\nonumber
&=&{\cal G}_{\sigma,+}^{\rm band}(\omega)
+  {\cal G}_{\sigma,-}^{\rm band}(\omega)
+\frac{\mod{m_0}/2}{\omega-\I\eta}
+\frac{\mod{m_0}/2}{\omega+\I\eta}\\
&\;&
+\frac{\I\pi}{2}\mod{m_0}\left(\delta(\omega-0^-)-\delta(\omega-0^+)\right)
\: ,
\label{gbandrearr}
\end{eqnarray}
where ${\cal G}_{\sigma,\pm}^{\rm band}(\omega)$ 
are the retarded/advanced components
of ${\cal G}_{\sigma}^{\rm band}(\omega)$. 
Inserting~(\ref{gbandrearr}) into~(\ref{piipt}), and
performing a simple contour integration, yields for $\Pi(\omega)$
\begin{eqnarray}
\label{pirearr}
\nonumber
\Pi(\omega)&=&\frac{(\mod{m_0}/2)^2}{\omega-\I\eta}
-\frac{(\mod{m_0}/2)^2}{\omega+\I\eta}\\
&+&\mod{m_0}\left\{{\cal G}_{\sigma,-}^{\rm band}(\omega)
                  -{\cal G}_{\sigma,+}^{\rm band}(\omega)
\right\}+I_1(\omega)\: ,
\end{eqnarray}
where 
\begin{equation}
I_1(\omega)=-
\int_{-\infty}^{\infty} \frac{\diff\omega_1}{2\pi\I}
{\cal G}_{\sigma}^{\rm band}(\omega_1)
{\cal G}_{\sigma}^{\rm band}(\omega_1-\omega)\; .
\end{equation}
Note that $\Im\Pi(\omega)$ has a pole contribution
at $\omega=0$, specifically
\begin{equation}
\label{iptpidelta}
\Pi(\omega)=\frac{\I\pi}{2}\mod{m_0}^2
\delta(\omega)+\Pi^{\rm band}(\omega)\; ,
\end{equation}
with
\begin{equation}
\Pi^{\rm band}(\omega)=\mod{m_0}\left\{
{\cal G}_{\sigma,-}^{\rm band}(\omega)-
{\cal G}_{\sigma,+}^{\rm band}(\omega)
\right\}+I_1(\omega)\; .
\end{equation}
Insertion of~(\ref{gbandrearr}) and~(\ref{pirearr}) 
into~(\ref{sipt}) yields for the self-energy
\begin{eqnarray}
\label{siptfull}
\Sigma_{\sigma}(\omega)&=&
U^2\left(\frac{\mod{m_0}}{2}\right)^2\left\{3{\cal G}^{\rm band}(\omega)
+\frac{\mod{m_0}}{\omega+\I\eta\, \sgn(\omega)}\right\}
\nonumber \\
&+&
I_2(\omega)+\frac{\mod{m_0}}{2}
U^2\left\{I_1^-(\omega)-I_1^+(\omega)\right\}\; ,
\end{eqnarray}
where 
\begin{equation}
I_2(\omega)=U^2 \int_{-\infty}^{\infty} \frac{\diff\omega_1}{2\pi \I}
{\cal G}_{\sigma}^{\rm band}(\omega-\omega_1)
\Pi^{\rm band}(\omega_1)
\end{equation}
and
\begin{equation}
I_1^{\pm}(\omega)=\pm\frac{1}{\pi}\int_{-\infty}^{\infty}
\frac{\diff\omega_1\Im I_1 
(\omega_1)}{\omega_1-\omega\mp \I\eta}
\; .
\end{equation}
$\Im I_1(\omega)$ may be found using a convolution involving only
$\Im{\cal G}_{\sigma}^{\rm band}(\omega)$,
and likewise $\Im I_2(\omega)$ may be found from a convolution of
$\Im{\cal G}_{\sigma}^{\rm band}(\omega)$ 
and $\Im\Pi^{\rm band}(\omega)$. For $\omega>0$,
\begin{eqnarray}
\label{convol1}
\Im I_1(\omega)&=&
\frac{1}{\pi}\int_{0}^{\omega} \diff\omega_1
\Im{\cal G}_{\sigma}^{\rm band}(\omega_1)
\Im{\cal G}_{\sigma}^{\rm band}(\omega-\omega_1)\; , \\
\Im I_2(\omega)&=&
\frac{U^2}{\pi}\int_0^{\omega} \diff\omega_1
\Im{\cal G}_{\sigma}^{\rm band}(\omega_1)
\Im{\Pi}^{\rm band}(\omega-\omega_1)\; .
\nonumber \\
&&
\label{convol2}
\end{eqnarray}
In order to calculate $\Im\Sigma_{\sigma}(\omega)$ 
for $\omega>0$, it is thus necessary, see~(\ref{siptfull}), 
to perform only the two
convolutions~(\ref{convol1}) and (\ref{convol2}), 
$\Im\Sigma_{\sigma}(\omega)$ for
$\omega<0$ follows immediately by symmetry, whence
$\Re\Sigma_{\sigma}(\omega)$ follows by Hilbert transform.

The above is what we implement numerically. 
However, when $U\to\infty$
eq.~(\ref{siptfull}) simplifies. In this limit~(\ref{muU})
gives $\mod{m_0}\to 1$; note that $\mod{m_0}\to 0$ 
is not a
self-consistent possibility, since $\mod{m_0}\ge 2/3$, cf.~(\ref{mumin}).
Specifically, 
\begin{equation}
\label{m0Uinf}
\mod{m_0}=1-\frac{4}{U^2}\quad, \quad  U\to\infty \; .
\end{equation}
Since $\mod{m_0}$ is the weight of the pole in 
${\cal G}_{\sigma}(\omega)$, (\ref{gband}), and
$\int_{-\infty}^{\infty}
\mod{\Im{\cal G}_{\sigma}(\omega)}\diff\omega=\pi$, we have
\begin{equation}
\Im{\cal G}_{\sigma}^{\rm band}(\omega)
\sim{\cal O}(U^{-2})\quad,\quad U\to\infty\; .
\end{equation}
Now, eqs.~(\ref{convol1}) and (\ref{convol2}) result in
$I_1(\omega)\sim{\cal O}(U^{-4})$ and
$I_2(\omega)\sim{\cal O}(U^{-2})$, whereby~(\ref{siptfull}) 
reduces to 
\begin{equation}
\Sigma_{\sigma}(\omega)=U^2\left(\frac{\mod{m_0}}{2}\right)^2
\left\{3{\cal G}_{\sigma}^{\rm band}(\omega)
+\frac{\mod{m_0}}{\omega+\I\eta\, \sgn(\omega)}\right\}\; ,
\end{equation}
with corrections ${\cal O}(U^{-2})$.
Using~(\ref{gband}) and ~(\ref{scrgipt}) this
may be rewritten as
\begin{equation}
\label{siptsc}
\Sigma_{\sigma}(\omega)
=\frac{-3\mod{m_0}^2U^2/4}{U/2-\omega'+G_{\sigma}(\omega)}
+\frac{U^2\mod{m_0}^3/2}{U/2-\omega'}+{\cal O}(U^{-2})\; ,
\end{equation}
where $\omega'=\omega+U/2$.
In the region of the lower Hubbard band only
($\omega'\sim{\cal O}(1)$) eq.~(\ref{siptsc})
may be expanded to give
\begin{eqnarray}
\Sigma_{\sigma}(\omega)&=&
-\frac{U}{2}-\omega'+3G_{\sigma}(\omega)
\label{siptlhb}
\\
&&-\frac{2}{U}\left[\omega'^2-6\omega'G_{\sigma}(\omega)
+3G_{\sigma}(\omega)^2\right]
+{\cal O}(U^{-2})\; .
\nonumber
\end{eqnarray}
If self-energy terms of ${\cal O}(U^{-1})$ and above are ignored, then
$G_{\sigma}(\omega)$ 
in the region of the lower Hubbard band reduces to
\begin{equation}
G_{\sigma}(\omega)=\frac{1/2}{\omega'
-2G_{\sigma}(\omega)}\; ,
\end{equation}
and the single-particle density of states becomes
\begin{equation}
D_{{\rm LHB},\sigma}(\omega)=\frac{1}{4\pi}
\sqrt{4-\omega'^2}
\; ,
\end{equation}
i.e., the correct strong coupling spectrum of a semi-ellipse with the full
noninteracting band-width. Retaining the ${\cal O}(U^{-1})$ terms in eq.~(\ref{siptlhb})
gives the ${\cal O}(U^{-1})$ corrections to the position and shape of the LHB.
The resultant spectrum is a distorted semi-ellipse of the form
\begin{eqnarray}
D_{{\rm LHB},\sigma}(\omega)&=&
\sqrt{4-\left(\omega'-\frac{a_{\rm IPT}}{U}\right)^2+{\cal O}(U^{-2})}
\nonumber \\
&& \times \frac{1-(a_{\rm IPT}/U)\omega'+{\cal O}(U^{-2})}{4\pi}\; ,
\label{diptsc}
\end{eqnarray}
where $a_{\rm IPT}=5/4$ 
and the LHB has band edges at $\omega'=(a_{\rm IPT}/U)\pm 2$. 
The IPT result
may be compared with the exact result to this order which also has the form of 
eq.~(\ref{diptsc}) but with a different coefficient of $a=1/2$.

\section{Local Moment Approach (LMA) in strong coupling}
\label{AppendixB}

The LMA of Logan et al.~is described in detail in Ref.~\cite{LMA},
where it is proved that the LMA recovers the strong coupling
single-particle spectrum up to and including terms of 
${\cal O}(U^0)$. Unlike IPT, the LMA also recovers the strong coupling
spectrum in the antiferromagnetic phase~\cite{LMA,LMAPRL}. 
The success of the LMA in strong coupling is unsurprising
since the LMA is explicitly motivated by ideas of hole
motion in a spin background. In this appendix we extend the previous
analysis to recover the 
${\cal O}(U^{-1})$ corrections to the LMA self-energy and single-particle
spectrum in the strong-coupling limit. This is done in three stages. First
the necessary equations of the LMA are given. Secondly, it is shown that
only the coupling of hole motion to zero-frequency spin-flips survive, with
the coupling of hole motion to higher energy particle-hole  excitations  only
entering at ${\cal O}(U^{-2})$. Finally, the resultant self-energy is 
expanded to ${\cal O}(U^{-1})$ and the corresponding 
density of states is obtained.

The bare propagators within the LMA are 
unrestricted Hartree Fock (UHF) 
propagators. Lattice sites are taken to have up-spin or down-spin 
permanent local moments at random, and labeled `A-type' or `B-type' 
respectively. We focus solely on an A-type site (Green functions for
B-type sites follow by symmetry) and suppress the `A' labels in what follows.
For the $Z=\infty$ Bethe lattice the UHF equations are 
simply 
\begin{equation}
G_{\sigma}^0(\omega)=-G_{-\sigma}^0(-\omega)
=[\omega+\frac{\sigma}{2}U\mod{\mu}-G^0(\omega)]^{-1} \; ,
\end{equation} 
where $\sigma=\pm1$ for $\uparrow/\downarrow$ respectively,  
and the symmetric single-particle 
Green function obeys $G^0(\omega)=[G_{\uparrow}^0(\omega)+
G_{\downarrow}^0(\omega)]/2$. 
The local moment is derived from
\begin{eqnarray}
\mu&=&\int_{-\infty}^0 \diff \omega
[D^0_{\uparrow}(\omega)-D^0_{\downarrow}(\omega)]
\; , \\
D^0_{\sigma}(\omega)&=&
-\frac{1}{\pi}\sgn(\omega)\Im G^0_{\sigma}(\omega)\; .
\end{eqnarray}
The RPA on-site particle-hole propagator 
\begin{equation}
\label{eq:lmapi}
\Pi^{+-}(\omega)=\frac{\Pi^{+-}_0(\omega)}{1-U\Pi^{+-}_0(\omega)},
\end{equation}
with
\begin{equation}
\label{eq:lmapi0}
\Pi^{+-}_0(\omega)=-\int_{-\infty}^{\infty}\frac{\diff\Omega}{2\pi \I}
G_{\downarrow}^0(\Omega)
G_{\uparrow}^0(\Omega-\omega)\; ,
\end{equation}
has a zero-frequency pole~\cite{LMA} (as well as higher-energy Stoner
bands) reflecting the presence of zero-frequency spin-flip excitations in 
the system. In the full LMA single-particle Green function,
\begin{equation}
\label{eq:lmagaup}
G_{\sigma}(\omega)=\left[\omega+\frac{\sigma}{2}U\mod{\mu}-G(\omega)
-\Sigma_{\sigma}(\omega)\right]^{-1},
\end{equation}
hole motion and accompanying spin-flips are dynamically coupled
via the self-energy
\begin{equation}
\label{eq:lmasigmafull}
\Sigma_{\uparrow}(\omega)=-\Sigma_{\downarrow}(-\omega)=
U^2\int\frac{\diff\Omega}{2\pi \I}\Pi^{+-}(\Omega)
{\cal G}_{\downarrow}(\omega+\Omega).
\end{equation}
The LMA equations must be solved self-consistently, since the
host Green function reads
\begin{equation}
{\cal G}_{\sigma}(\omega)=
[\omega+\frac{\sigma}{2}U\mod{\mu}-G(\omega)]^{-1}
\end{equation}
with $G(\omega)=[G_{\uparrow}(\omega)+G_{\downarrow}(\omega)]/2$.

To calculate the LMA self-energy to  ${\cal O}(U^{-1})$, we proceed 
in two stages. First we sketch the proof that only the zero-frequency spin-flip
pole in $\Pi^{+-}(\omega)$ need be retained and the self-energy simplifies to
\begin{eqnarray}
\label{eq:lmasigmapole}
\Sigma_{\uparrow}(\omega)&=&
U^2{\cal G}_{\downarrow}^-(\omega)
+{\cal O}(U^{-2})\\
&=&U^2\int_{-\infty}^0\frac{\diff\omega_1}{\pi}
\frac{\Im{\cal G}_{\downarrow}(\omega_1)}{\omega-\omega_1-\I\eta}
+{\cal O}(U^{-2})\; .
\label{eq:lmaht}
\end{eqnarray}
Subsequently we analyze ${\cal G}_{\downarrow}^-(\omega)$ 
to obtain the spectrum to ${\cal O}(U^{-1})$. 

It has been shown \cite{LMA} that the RPA particle-hole propagator may be
written
\begin{equation}
\label{eq:lmaQ}
\Pi^{+-}(\omega)=\frac{Q}{-\omega-\I\eta}+\Pi^{+-}_>(\omega)
+\Pi^{+-}_<(\omega)\; ,
\end{equation}
where $\Pi^{+-}_>(\omega)$ and $\Pi^{+-}_<(\omega)$ 
are the Stoner-like contributions whose imaginary
parts are bands centered at $\pm U$ respectively. 
The full self-energy is thus
\begin{equation}
\Sigma_{\uparrow}(\omega)=QU^2{\cal G}_{\downarrow}^-(\omega)
+\Sigma^{\rm S}_{\uparrow}(\omega)\; ,
\end{equation}
where
\begin{equation}
\label{eq:lmaimsstoner}
\Im\Sigma^{\rm S}_{\uparrow}(\omega)
=\left\{ \begin{array}{@{}lcl@{}}
{ -U^2 \int_0^{\omega}\diff\Omega{\cal D}_{\downarrow}(\Omega)
\Im \Pi_<^{+-}(\Omega-\omega)} & \hbox{for} & \omega>0 \\[6pt]
{ U^2\int_{\omega}^{0}\diff\Omega{\cal D}_{\downarrow}(\Omega)
\Im \Pi_>^{+-}(\Omega-\omega)} & \hbox{for} & \omega<0
\end{array}\right. \, .
\end{equation}
To show that $\Sigma^{\rm S}_{\uparrow}(\omega)\sim{\cal O}(U^{-2})$, it is 
necessary to examine $\Pi^{+-}(\omega)$ 
and hence $\Pi^{+-}_0(\omega)$ and $G_{\sigma}^0(\omega)$. 
By expanding $G_{\downarrow}^0(\omega)$ in the region of the LHB
it quickly follows that 
\begin{equation}
\mod{\mu}=1-1/U^2+{\cal O}(U^{-3})\; .
\end{equation}
Then $\Im \Pi^{+-}_0(\omega)$
may be shown from~(\ref{eq:lmapi0}) to
have a minority band centered around $\omega\approx -U$ with spectral weight
$\left[(1-\mod{\mu})/2\right]^2=1/(4U^4)+{\cal O}(U^{-5})$, and a
majority band centered at $\omega\approx U$ with spectral weight 
$\left[(1+\mod{\mu})/2\right]^2=1-1/(U^2)+{\cal O}(U^{-3})$. 
The bands of the RPA propagator which follow from~(\ref{eq:lmapi}),
\begin{equation}
\Im\Pi^{+-}(\omega)=
\frac{\Im\Pi^{+-}_0(\omega)}{\mod{1-U\Pi^{+-}_0(\omega)}^2}\; ,
\end{equation} 
are also centered
around $\omega=\pm U$. Since $\Pi^{+-}_0(\omega)\sim{\cal O}(1)$ in the
region of the upper band it follows that the
spectral weight in $\Im\Pi^{+-}_>(\omega)$ is ${\cal O}(U^{-2})$. 
In the region of the lower Hubbard band ($\omega\approx -U/2$) 
Kramers-Kronig relations result in 
$\Re\Pi^{+-}_0(\omega)=1/(2U)+{\cal O}(U^{-2})$, 
whence $\Im \Pi^{+-}(\omega)=4\Im \Pi^{+-}_0(\omega)
+{\cal O}(U^{-5})$ and the spectral weight of $\Im \Pi^{+-}_<(\omega)$ 
is $1/U^4$.
The remainder of the spectral weight of $\Im \Pi^{+-}(\omega)$ 
is contained in 
the pole weight $Q$. Eqs.~(\ref{eq:lmapi}), (\ref{eq:lmapi0}) 
may be used to show
that 
\begin{equation}
\Pi^{+-}(\omega\to\infty)=\Pi^{+-}_0(\omega\to\infty)
=-\frac{\mod{\mu}}{\omega}\; , 
\end{equation}
whence Kramers-Kronig relations give the missing 
weight as $Q=1-{\cal O}(U^{-2})$. Finally, 
since ${\cal G}_{\downarrow}(\omega)$
(in common with $G^0_{\downarrow}(\omega)$) 
has minority and majority spectral weights
that approach $1/(2U^2)$ and $1-1/(2U^2)$ respectively, 
eq.~(\ref{eq:lmaimsstoner}) shows
the contributions to $\Im\Sigma_{\uparrow}^{\rm S}(\omega)$ 
are ${\cal O}(U^{-2})$.
A more detailed analysis shows that in the region 
of the LHB $\Sigma_{\uparrow}^{\rm S}(\omega)$
is pure real and ${\cal O}(U^{-4})$, and thus 
neglecting $\Sigma_{\uparrow}^{\rm S}(\omega)$
barely changes the LMA results throughout the insulating phase. 
Nevertheless the analysis outlined here is sufficient 
to arrive at~(\ref{eq:lmasigmapole}).

Substituting~(\ref{eq:lmasigmapole}) together with $\mod{\mu}=1-1/U^2+
{\cal O}(U^{-3})$
into~(\ref{eq:lmagaup}) yields
in the region of the LHB
\begin{equation}
\label{eq:lmagupstage1}
G_{\uparrow}(\omega)=\left[\omega'-\half G_{\uparrow}(\omega)
-U^2{\cal G}_{\downarrow}^-(\omega)+{\cal O}(U^{-2})\right]^{-1}\; ,
\end{equation}
where we have defined 
\begin{equation}
\omega'=\omega+U/2\; ,
\end{equation}
and we have also used the LHB expansion
\begin{equation}
\label{eq:lmaGexp}
G(\omega)=\frac{1}{2} G_{\uparrow}(\omega)-\frac{1}{2U}+{\cal O}(U^{-2})\; .
\end{equation}
All that remains is to calculate ${\cal G}_{\downarrow}^-(\omega)$ to 
${\cal O}(U^{-3})$. We focus exclusively on the LHB.
First we expand ${\cal G}_{\downarrow}(\omega)$
in powers of $1/U$,
\begin{eqnarray}
\nonumber
{\cal G}_{\downarrow}(\omega)=&-&\frac{1}{U}-\frac{1}{U^2}
\left[\omega'+\frac{1}{2U}-G(\omega)\right]\\
&-&\frac{1}{U^3}\left[\omega'
+\frac{1}{2U}-G(\omega)\right]^2+{\cal O}(U^{-4})\; .
\end{eqnarray}
Since in the region of the LHB 
$\Im{\cal G}_{\downarrow}^-(\omega)=\Im{\cal G}_{\downarrow}(\omega)$
by definition, we have
\begin{eqnarray}
\nonumber
U^2\Im {\cal G}_{\downarrow}^-(\omega)
&=&\Im G(\omega)-\frac{1}{U}\Im\left\{[\omega'-G(\omega)]^2
\right\}+{\cal O}(U^{-2})\\
&=&\half\Im G_{\uparrow}(\omega)-\frac{3}{4U}\Im\left\{\omega'G_{\uparrow}(\omega)\right\}
+{\cal O}(U^{-2})\; ,
\nonumber \\
&&
\end{eqnarray}
where we have used~(\ref{eq:lmaGexp}) and the exact form of the 
single-particle Green function in strong coupling,
\begin{equation}
G_{\uparrow}^2(\omega)
=\omega'G_{\uparrow}(\omega)-1+{\cal O}(U^{-1})\; , 
\end{equation}
which is correctly obtained within the LMA. 
The Hilbert transform~(\ref{eq:lmaht}) then
yields
\begin{eqnarray}
\nonumber
U^2 {\cal G}_{\downarrow}^-(\omega)&=&\frac{1}{2}G_{\uparrow}(\omega)
+\frac{3\omega'}{4U}\I\Im G_{\uparrow}(\omega)\\
\nonumber
&~&+\frac{3}{4U}{\rm P}\!\int_{\rm LHB}\diff\omega_1
\frac{\omega_1 D_{\uparrow}(\omega_1)}
{\omega-\omega_1}+{\cal O}(U^{-2})\\
&=&\frac{1}{2}G_{\uparrow}(\omega)+\frac{3\omega'}{4U}G_{\uparrow}(\omega)
-\frac{3}{4U}+{\cal O}(U^{-2})\; ,\nonumber
\\
&& \label{eq:lmagscrsimple}
\end{eqnarray}
where we have used $2\pi D_{\uparrow}(\omega)
=\sqrt{4-\omega'^2+{\cal O}(U^{-1})}$
to simplify the ${\cal O}(U^{-1})$ correction. Eq.~(\ref{eq:lmagscrsimple})
is only valid in the region of LHB spectral density. Substituting 
(\ref{eq:lmagscrsimple}) into~(\ref{eq:lmagupstage1}) gives the 
single-particle Green function in this region,
\begin{equation}
G_{\uparrow}(\omega)=\left[\omega'+\frac{3}{4U}-\left\{1+\frac{3\omega'}{4U}
\right\}G_{\uparrow}(\omega)+{\cal O}(U^{-2})\right]^{-1}\, .
\end{equation}
The corresponding single-particle spectrum in strong coupling 
including $1/U$ corrections is
\begin{eqnarray}
D_{\uparrow}(\omega)&=&
\sqrt{4-\left(\omega'-\frac{a_{\rm LMA}}{U}\right)^2
+{\cal O}(U^{-2})}
\nonumber \\
 && \times
\frac{1-(a_{\rm LMA}/U)\omega' +{\cal O}(U^{-2}) }{2\pi}
\end{eqnarray}
with the LMA coefficient $a_{\rm LMA}=3/4$ instead of $a=1/2$
from the strong-coupling expansion.


\begin{thebibliography}{99}
%
\bibitem{Mottbook} N.F.~Mott, \textit{Metal--Insulator Transitions},
2nd edition (Taylor and Francis, London, 1990).
%
\bibitem{Gebhardbook} F.~Gebhard, \textit{The Mott Metal--Insulator Transition}
(Sprin\-ger, Berlin, 1997).
%
\bibitem{RMP} A.~Georges, G.~Kotliar, W.~Krauth, and
M.J.~Rozenberg, Rev.\ Mod.\ Phys.~{\bf 68}, (1996) 13.
%
\bibitem{Imada} M.\ Imada, A.\ Fujimori, and Y.\ Tokura,
Rev.\ Mod.\ Phys.\ {\bf 70}, (1998) 1039.
%
\bibitem{LiebWu} E.H.~Lieb and F.Y.~Wu, Phys.~Rev.~Lett.~{\bf 20}, (1968) 1445.
%
\bibitem{GebhardRuckenstein} F.~Gebhard and A.E.~Ruckenstein, 
Phys.~Rev.~Lett.~{\bf 68}, (1992) 244; 
F.~Gebhard, A.~Girndt, and A.E.~Ruckenstein,
Phys.~Rev.~B~{\bf 49}, (1994) 10926.
%
\bibitem{MV} W.~Metzner and D.~Vollhardt, Phys.~Rev.~Lett.~{\bf 62},
(1989) 324.
%
\bibitem{MHa} E.~M\"uller-Hartmann, Z.~Phys.~B~{\bf 74}, 507 (1989).
%
\bibitem{BrandtMielsch} U.~Brandt and C.~Mielsch, Z.~Phys.~B~{\bf 75},
(1989) 365, {\sl ibid.}~{\bf 79}, (1990) 295.
%
\bibitem{Jarrell} M.~Jarrell, Phys.~Rev.~Lett.~{\bf 69}, (1992) 168.
%
\bibitem{RDA} R.M.~Noack and F.~Gebhard,
Phys.~Rev.~Lett.~{\bf 82}, (1999) 1915.
%
\bibitem{7Schwaben} J.~Schlipf, M.~Jarrell, P.G.J.~van Dongen, 
N.~Bl\"umer, S.\ Kehrein, T.\ Pruschke, and D.\ Vollhardt,
Phys.\ Rev.\ Lett.~{\bf 82}, (1999) 4890.
%
\bibitem{7Schwabenreply} M.J.~Rozenberg, R.~Chitra, and G.~Kotliar,
Phys.\ Rev.\ Lett.~{\bf 83}, (1999) 3498.
%
\bibitem{Krauth} W.~Krauth, Phys.~Rev.~B~{\bf 62}, (2000) 6860.
%
\bibitem{BullaPRL} R.~Bulla, Phys.~Rev.~Lett.~{\bf 83}, (1999) 136. 
%
\bibitem{BullaVolli} R.~Bulla, T.A.~Costi, and D.~Vollhardt,
Phys.~Rev.~B~{\bf 64}, 045103 (2001).
%
\bibitem{BullaLDMFT} R.~Bulla and M.~Potthoff, 
Eur.~Phys.~J.~B~{\bf 13}, (2000) 257.
%
\bibitem{Ono} Y.~\={O}no, R.~Bulla, A.C.~Hewson,
and M.~Potthoff, Eur.\ Phys.\ J.~B~{\bf 22}, (2001) 283. 
%
\bibitem{KotliarFisher} G.~Moeller, Q.~Si, G.~Kotliar, 
M.J.~Rozenberg, and D.S.\ Fisher, 
Phys.\ Rev.\ Lett.~{\bf 74}, (1995) 2082.
%
\bibitem{LoganNozieres} D.E.~Logan and P.~Nozi\`eres,
Phil.~Trans.~R.~Soc.\ London A~{\bf 356}, (1998) 249;
P.~Nozi\`eres, Eur.~Phys.~J.~B~{\bf 6}, (1998) 447.
%
\bibitem{Kehrein} S.~Kehrein, Phys.~Rev.~Lett.~{\bf 81}, (1998) 3912.
%
\bibitem{Kotliarreply} A.~Georges and G.~Kotliar, 
Phys.~Rev.~Lett.~{\bf 84}, (2000) 3500;
S.~Kehrein, Phys.~Rev.~Lett.~{\bf 84}, (2000) 3501.
%
\bibitem{KotliarIPT} M.J.~Rozenberg, G.~Kotliar, and X.Y.~Zhang,
Phys.\ Rev.\ B~{\bf 49}, (1994) 10181.
%
\bibitem{NCA} T.~Pruschke, D.L.~Cox, and M.~Jarrell, 
Europhys.~Lett.~{\bf 21}, (1993) 593;
Phys.~Rev.~B~{\bf 47}, (1993) 3553;
M.~Jarrell, J.K.~Freericks, and T.~Pruschke, Phys.~Rev.~B~{\bf 51},
(1995) 11704.
%
\bibitem{LMA} D.E.~Logan, M.P.~Eastwood, and M.A.~Tusch,
J.\ Phys.\ Cond.\ Matt.~{\bf 9} (1997) 4211.
%
\bibitem{CK} M.~Caffarel and W.~Krauth,
Phys.~Rev.~Lett.~{\bf 72}, (1994) 1545.
%
\bibitem{Rozenbergetal} M.J.~Rozenberg, G.~M\"oller, and G.~Kotliar,
Mod.~Phys.\ Lett.\ B~{\bf 8} (1994), 535.
%
\bibitem{andere} See also 
Q.~Si, M.J.~Rozenberg, G.~Kotliar and A.E.~Ruckenstein,
Phys.~Rev.~Lett.~{\bf 72}, (1994) 2761;
M.J.~Rozenberg, G.~Kotliar, H.~Kaj\"uter, G.A.~Thomas, D.H.~Rapkine,
J.M.~Honig, and P.~Metcalf, Phys.~Rev.~Lett.~{\bf 75}, (1995) 105.
%
\bibitem{Eva} E.~Kalinowski and F.~Gebhard, J.~Low Temp.~Phys.~{\bf 126},
(2002) 979. 
%
\bibitem{PvDetal} P.G.J.~van Dongen, F.~Gebhard, and D.~Vollhardt,
Z.\ Phys.\ B~{\bf 76}, (1989) 199.
%
\bibitem{Economou} E.~Economou, \textit{Green's Functions in Quantum Physics},
2nd edition (Springer, Berlin, 1983).
%
\bibitem{HubbardI} J.~Hubbard, Proc.~Roy.~Soc.\ London
Ser.~A~{\bf 276}, (1963) 238; {\sl ibid.}~{\bf 277}, (1963) 237.
%
\bibitem{Fetter} A.L.~Fetter and J.D.~Walecka, 
\textit{Quantum Theory of Many-Particle Systems} 
(McGraw--Hill, New York, 1971). 
%
\bibitem{Kato} T.~Kato, Prog.~Theor.~Phys.~{\bf 4}, (1949) 154;
A.~Messiah, \textit{Quantum Mechanics}, Vol.~2 
(North-Holland, Amsterdam, 1962) \S~16.
%
\bibitem{Takahashi} M.~Takahashi, J.~Phys.~C~{\bf 10}, (1977) 1289.
%
\bibitem{Anderson} P.W.~Anderson, Phys.~Rev.~{\bf 115}, (1959) 2.
%
\bibitem{MVSchmit} W.~Metzner, P.~Schmit, and D.~Vollhardt,
Phys.~Rev.~B~{\bf 45}, (1992) 2237;  
W.F.~Brinkman and T.M.~Rice, Phys.\ Rev.\ B~{\bf 2}, (1970) 1324.
%
\bibitem{PvD} P.G.J.\ van Dongen and D.~Vollhardt,
Phys.~Rev.~Lett.~{\bf 65}, (1990) 1663;
P.G.J.\ van Dongen, Phys.~Rev.~B~{\bf 45}, (1992) 2267.
%
\bibitem{Potthoff}  M.~Potthoff, preprint cond-mat/0301137 (2003),
unpublished.
%
\bibitem{iptoriginal}  A.~Georges and G.~Kotliar,
Phys.\ Rev.\ B~{\bf 45}, (1992) 6479.
%
\bibitem{Mikethesis} M.P.~Eastwood, Ph.D.~thesis, Oxford University
(1998), unpublished.
%
\bibitem{HubbardIII} J.~Hubbard, Proc.~Roy.~Soc.\ London
{\bf 281}, (1964) 401.
%
\bibitem{LMAPRL} D.E.~Logan, M.P.~Eastwood, and M.A.~Tusch,
Phys.\ Rev.\ Lett.~{\bf 76}, (1997) 4785.
%
\end{thebibliography}
\end{document}